\documentclass[fleqn,usenatbib]{mnras}

\usepackage{newtxtext,newtxmath}

\usepackage[T1]{fontenc}

\DeclareRobustCommand{\VAN}[3]{#2}
\let\VANthebibliography\thebibliography
\def\thebibliography{\DeclareRobustCommand{\VAN}[3]{##3}\VANthebibliography}

\newcommand{\curvefit}{{\fontfamily{pcr}\selectfont curve\_fit}\,}
\newcommand{\spright}{{\fontfamily{pcr}\selectfont spright}\,}

\usepackage{graphicx}	
\usepackage{amsmath}	

\usepackage{graphicx}	
\usepackage{amsmath}	
\usepackage{xspace}
\usepackage{longtable}
\usepackage{subfig}
\usepackage{booktabs}
\usepackage{lscape}
\usepackage{siunitx}
\usepackage[symbol]{footmisc}
\usepackage[dvipsnames]{xcolor}

\usepackage{tabularx}

\title[TOI-7166\,b system]{TOI-7166\,b: A Habitable Zone mini-Neptune planet around a nearby low-mass star}

\author[K. Barkaoui et al.]
{
Khalid Barkaoui$^{\href{https://orcid.org/0000-0003-1464-9276}{\includegraphics[scale=0.5]{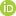}}}$,$^{1,2,3}$\thanks{E-mail: khalid.barkaoui@iac.es}
Francisco J.~Pozuelos$^{\href{https://orcid.org/0000-0003-1572-7707}{\includegraphics[scale=0.5]{figures/orcid.jpg}}}$,$^{4}$ 
Benjamin V.~Rackham$^{\href{https://orcid.org/0000-0002-3627-1676}{\includegraphics[scale=0.5]{figures/orcid.jpg}}}$,$^{3,5}$ 
Adam J.~Burgasser,$^{6}$   
\newauthor
Amaury H.M.J.~Triaud,$^{7}$ 
Miquel~Serra-Ricart$^{\href{https://orcid.org/0000-0002-2394-0711}{\includegraphics[scale=0.5]{figures/orcid.jpg}}}$,$^{1,8,9}$ 
Mathilde~Timmermans,$^{7,2}$  
Selçuk.~Yal\c{c}{\i}nkaya,$^{10,11,12}$    
\newauthor
Abderahmane~Soubkiou,$^{2}$ 
Keivan~G.~Stassun,$^{13}$ 
Karen A.~Collins$^{\href{https://orcid.org/0000-0001-6588-9574}{\includegraphics[scale=0.5]{figures/orcid.jpg}}}$,$^{14}$ 
Pedro J. Amado,$^{4}$
\"Ozgur~Ba\c{s}t\"urk,$^{10, 11}$ 
\newauthor
Artem~Burdanov,$^{3}$ 
Yasmin T. Davis$^{\href{https://orcid.org/0009-0000-6625-137X}{\includegraphics[scale=0.5]{figures/orcid.jpg}}}$,$^{7}$ 
Julien de Wit,$^{3}$
Brice-Olivier Demory,$^{15}$ 
Sarah Deveny,$^{23, 21}$
\newauthor
Georgina Dransfield$^{\href{https://orcid.org/0000-0002-3937-630X}{\includegraphics[scale=0.5]{figures/orcid.jpg}}}$,$^{24,25,7}$
Elsa~Ducrot,$^{16, 17}$ 
Michaël Gillon,$^{2}$
Yilen Gómez Maqueo Chew,$^{18}$ 
\newauthor
Matthew~J.~Hooton,$^{19}$
Keith Horne$^{\href{https://orcid.org/0000-0003-1728-0304}{\includegraphics[scale=0.5]{figures/orcid.jpg}}}$,$^{20}$
Steve~B. Howell,$^{21}$ 
Clàudia Janó Muñoz,$^{19}$ 
Emmanuel Jehin,$^{22}$
\newauthor
John~M.~Jenkins,$^{21}$
Colin Littlefield,$^{23, 21}$ 
Eduardo L. Martín,$^{1}$
Prajwal Niraula$^{\href{https://orcid.org/0000-0002-8052-3893}{\includegraphics[scale=0.5]{figures/orcid.jpg}}}$,$^{3}$
Peter P. Pedersen,$^{19,26}$ 
\newauthor
Dedier Queloz,$^{19,26}$
Madison G.~Scott$^{\href{https://orcid.org/0009-0006-3846-4558}{\includegraphics[scale=0.5]{figures/orcid.jpg}}}$,$^{7}$
Ramotholo Sefako$^{\href{https://orcid.org/0000-0003-3904-6754}{\includegraphics[scale=0.5]{figures/orcid.jpg}}}$,$^{27}$
Avi~Shporer,$^{5}$
Christopher Stockdale$^{\href{https://orcid.org/0000-0003-2163-1437}{\includegraphics[scale=0.5]{figures/orcid.jpg}}}$,$^{28}$ 
\newauthor
Emma Softich,$^{6}$
Alfredo Sota$^{\href{https://orcid.org/0000-0002-9404-6952}{\includegraphics[scale=0.5]{figures/orcid.jpg}}}$,$^{4}$
Benjamin Tofflemire,$^{29}$ 
\"Ozlem~\c{S}im\c{s}ir$^{\href{https://orcid.org/0009-0007-8172-6602}{\includegraphics[scale=0.5]{figures/orcid.jpg}}}$,$^{12}$ 
Roberto Varas,$^{4}$
\newauthor
Francis Zong Lang,$^{15}$
Sebastián S.~Z\'u\~niga-Fern\'andez$^{\href{https://orcid.org/0000-0002-9350-830X}{\includegraphics[scale=0.5]{figures/orcid.jpg}}}$,$^{2}$
\\
Affiliations are listed at the end of the paper.
}

\date{Accepted XXX. Received YYY; in original form ZZZ}

\pubyear{\the\year{}}

\begin{document}
\label{firstpage}
\pagerange{\pageref{firstpage}--\pageref{lastpage}}
\maketitle

\begin{abstract}
We present the discovery and validation of TOI-7166\,b, a $2.01 \pm 0.05~R_\oplus$ planet orbiting a nearby low-mass star. 
We validated the planet by combining TESS and multi-color high-precision photometric observations from  ground-based telescopes, together with spectroscopic data, high-contrast imaging, archival images, and statistical arguments.
The host star is an M4-type dwarf at a distance of $\sim$35\,pc from the Sun. It has a mass and a radius of $M_\star = 0.190 \pm 0.004\,M_\odot$ and $R_\star=0.222 \pm0.005\,R_\odot$, respectively.
TOI-7166\,b has an orbital period of 12.9\,days, which places it close to the inner edge of the Habitable Zone of its host star, receiving an insolation flux of $S_p=1.07\pm0.08\,S_\oplus$  and an equilibrium temperature of $T_{\rm eq} = 249\pm5$\,K (assuming a null Bond albedo). 
The brightness of the host star makes TOI-7166 a suitable target for radial velocity follow-up to measure the planetary mass and bulk density.
Moreover, the physical parameters of the system including the infrared brightness ($K_{\rm mag} = 10.6$)  of the star and the planet-to-star radius ratio ($0.0823 \pm 0.0012$) make TOI-7166\,b an exquisite target for transmission spectroscopic observations with the JWST, to constrain the exoplanet atmospheric compositions.
\end{abstract}

\begin{keywords}
{exoplanets -- Planetary Systems, planets and satellites: detection --
Planetary Systems, stars: individual, low-mass, late-type}
\end{keywords}



\section{Introduction} \label{sec:intro}

Over the past two decades, exoplanet science has experienced an extraordinary expansion, with more than 6,000 confirmed planets reported to date\footnote{\url{https://exoplanetarchive.ipac.caltech.edu/index.html}}. This rapid growth has been largely driven by dedicated space missions, such as \emph{Kepler} \citep{borucki2010} and the \emph{Transiting Exoplanet Survey Satellite} (\emph{TESS}; \citealt{Ricker_2015JATIS_TESS}), as well as extensive ground-based surveys and radial velocity programs. These efforts have revealed an unexpected diversity of planetary systems and populations. Within all these planets, sub-Neptune-sized (1.5-4.0\,R$_{\oplus}$) are among the most common yet enigmatic planetary populations known to date, with no Solar System analogues. Their origin and internal structure remain poorly understood, lying at the transition between rocky super-Earths and gas-rich Neptunes \citep[e.g.][]{Rogers_2015ApJ,bean2021}. In particular, their bulk compositions, atmospheric retention, and formation pathways are key open questions in exoplanet science. Around M dwarfs, these planets are especially interesting: their transits produce large signals, their radial velocity amplitudes are enhanced by the low stellar masses, and their habitable zones are located at short orbital periods \citep{Kasting_1993Icar,Kopparapu_2013ApJ}. These factors make them prime targets for precise mass measurements and detailed atmospheric studies. Indeed, the James Webb Space Telescope (JWST) is investing considerable amount of time on these candidates such as GJ 3470\,b \citep{beatty2024}, GJ 1214\,b \citep{Schlawin2014}, K2-18\,b \citep{madu2023}, TOI-270\,d \citep{holmberg2024}, LHS 1140\,b \citep{damiano2024}, and L 98-59\,d \citep{gressier2024} among others; we refer the reader to \cite{madu_2025_review} for a recent review of JWST observations of sub-neptunes. Of particular interest is the case of K2-18b, its nature as a water-rich Hycean planet \citep{madu2023,hu2025} and the subsequent claim of a detection of dimethyl sulfide (DMS) as a potential biosignature in its atmosphere \citep{madu2025_dms}, has produced one of the hottest debates today, with a series of independent studies refuting both the hycean hypothesis \citep[see, e.g.,][]{shorttle2024,wogan2024,werlen2025} and the detection of any biomaker in its atmosphere \cite[see, e.g.,][]{luque2025,stevenson2025,schmidt2025,welbanks2025}.

All of these highlight the community's interest in this puzzling planetary population and underscore the importance of identifying well-suited sub-Neptunes orbiting nearby M dwarfs to build benchmark systems for testing formation, evolution, and habitability models, as well as to enable future atmospheric characterization with state-of-the-art facilities such as the JWST and/or upcoming Extremely Large Telescopes.

In this context, this study reports the discovery of TOI-7166\,b, a temperate sub-Neptune-sized planet orbiting an M4-type star identified by \emph{TESS}. We characterize its host star by combining spectral energy distribution (SED) fitting with low-resolution spectroscopy, and we validate its planetary nature using space- and ground-based photometry along with statistical arguments. TOI-7166\,b has a radius of $\sim$2.01\,R$_\oplus$ and orbits its host star every 12.9 days, placing it in the habitable zone and receiving an insolation of S$\sim$1.07\,S$_\oplus$. The combination of the planet’s properties and the brightness of its host star makes TOI-7166\,b an excellent candidate for precise mass determination and detailed atmospheric characterization, placing it among the most promising sub-Neptunes known to date for such studies. 

The paper is organized as follows. Section~\ref{sec2} describes the observations used in this study and the stellar characterization. Section~\ref{sec:planet_validation} presents the validation of the planetary signal, while Section~\ref{sec:glbal_fit} details the global photometric modeling. Additional planet searches and \emph{TESS} detection limits are discussed in Section~\ref{sec:planet_searches}. Finally, we outline the prospects for future follow-up observations in Section~\ref{sec:prospects} and summarize our conclusions in Section~\ref{sec:conclusion}.

\section{Observations and data analysis}
\label{sec2}
\subsection{TESS data} \label{sec:tess_photometry}
\begin{figure}
	\centering
     \includegraphics[scale=0.5]{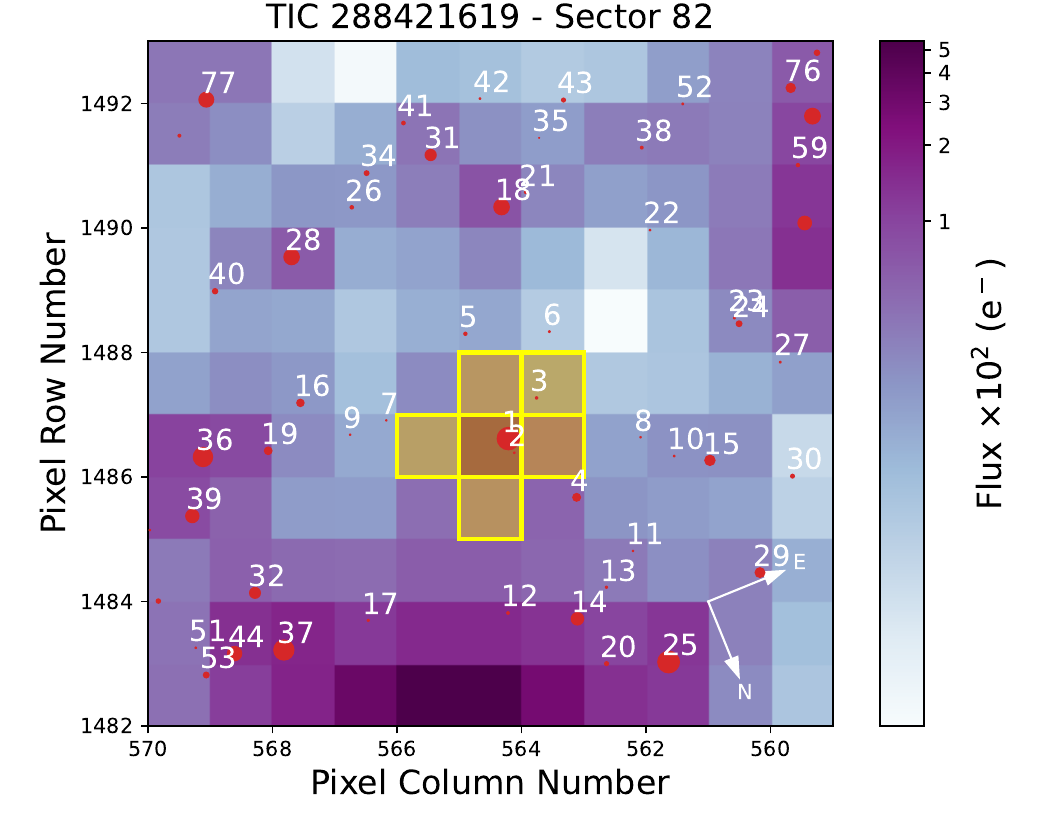} 
	\caption{\emph{TESS} target pixel file images of TOI-7166 observed in Sectors 82 made by \href{https://github.com/jlillo/tpfplotter}{\tt tpfplotter} \citep{Aller_2020AA}. Red dots show the location of Gaia DR3 sources, and the yellow shaded regions show the photometric apertures used for  photometric measurements extraction.} 
	\label{fig:Target_pixel}
\end{figure}

\begin{figure*}
	\centering
     \includegraphics[scale=0.36]{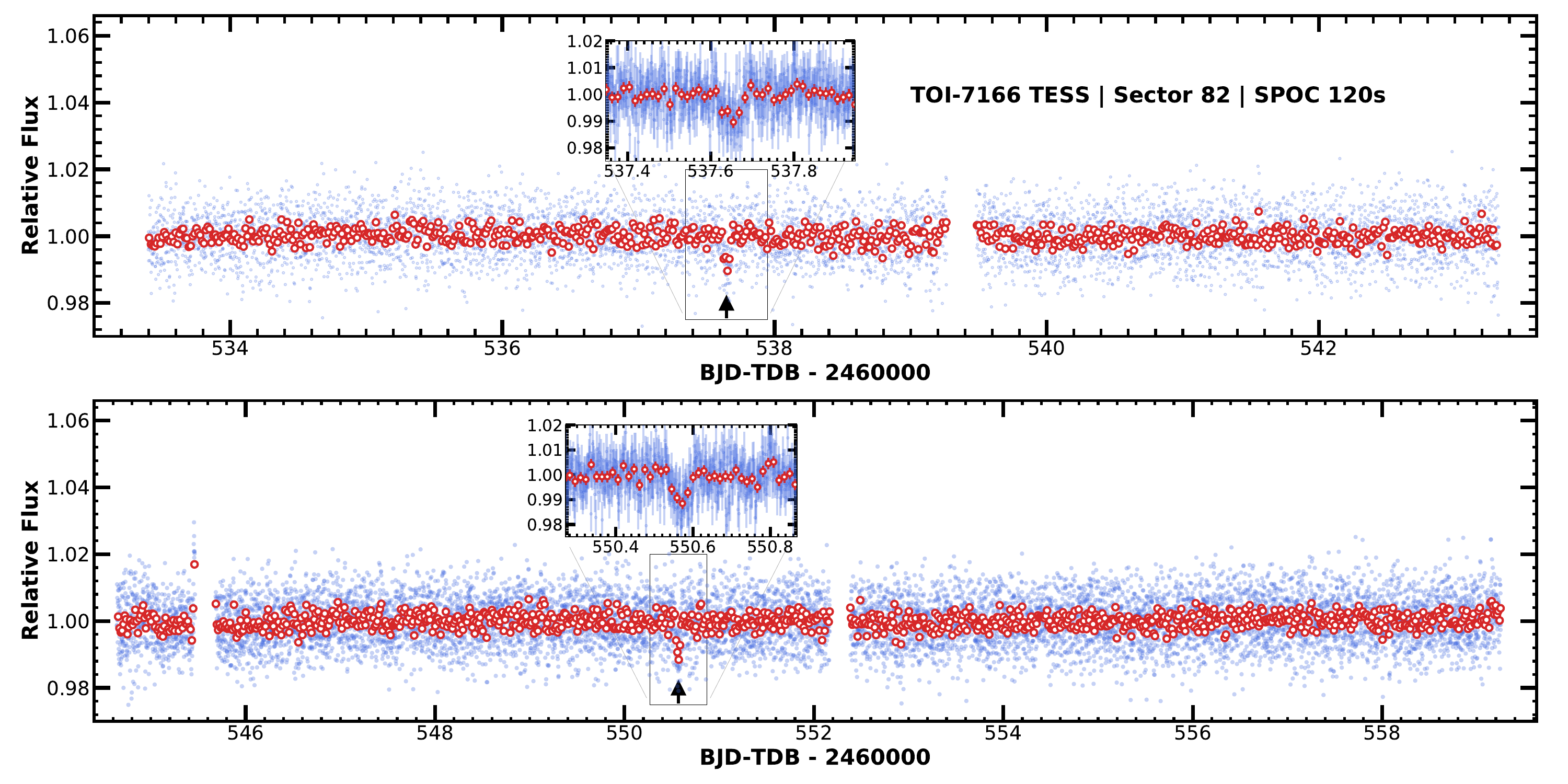} 
	\caption{\emph{TESS} PDC-SAP flux measurements extracted from the 2-min cadence data of TOI-7166. The target was observed in sector~82. The blue points show the 2-min data, and the red points show the 20-min binned data. The transit locations of TOI-7166\,b are shown with black arrows and zoom boxes. }
	\label{fig:TESS_LCs_time}
\end{figure*}

The host star TIC\,288421619 (TOI-7166) was observed by \emph{TESS} \citep{Ricker_2015JATIS_TESS} in Sector 82 for 27 days from August 10 to September 5, 2024 with 120-second and 200-second cadences, in Camera \#1 and CCD \#2. 
For our global analysis, we used the Pre-search Data Conditioning Simple Aperture Photometry flux (PDC-SAP; \citealt{Stumpe_2012PASP,Smith_2012PASP,Stumpe_2014}), constructed by the TESS Science Processing Operations Center (SPOC; \citealt{SPOC_Jenkins_2016SPIE}) at the Ames Research Center, from the Mikulski Archive for Space Telescopes\footnote{\url{https://archive.stsci.edu/missions-and-data/tess}}, as they are already calibrated for any instrument systematics and crowding effects. 
We extracted the normalized TOI-7166 fluxes using the {\tt lightkurve} \citep{Lightkurve_2018ascl} Python package. 
\autoref{fig:Target_pixel} shows the TESS FOV including the TESS aperture photometric and the location of nearby Gaia DR3 sources \citep{Gaia_Collaboration_2021A&A}. \autoref{fig:TESS_LCs_time} shows the TESS photometric data for TOI-7166. The transit events are highlighted by the black arrows and zoom boxes.

\subsection{Ground-based data} \label{sec:gb_photometry}

\begin{figure}
	\centering
     \includegraphics[scale=0.34]{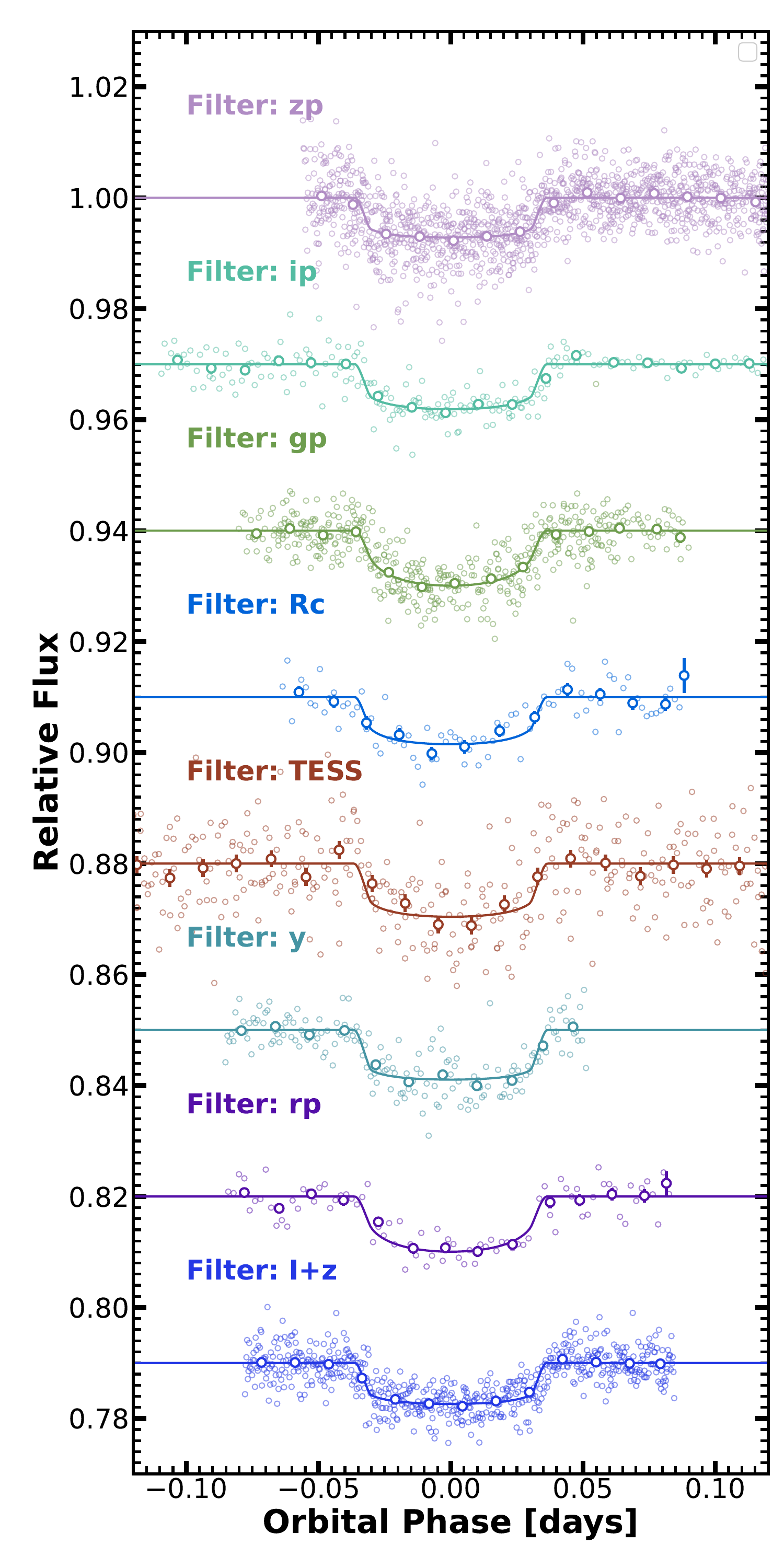} 
	\caption{\emph{TESS} and ground-based phase-folded transit light curves of TOI-7166\,b. The colored data points show the relative flux, and the colored solid lines show the best-fitting transit model superimposed. The transit light curves are shifted along y-axis for visibility.} 
	\label{fig:TESSGroundB_LCs_phased}
\end{figure}

All ground-based photometric time-series were scheduled based on the {\tt TESS Transit Finder} tool, which is a customized version of the {\tt Tapir} software package \citep{jensen2013}. These are summarized in the following sections. Figures\,\ref{fig:TESSGroundB_LCs_phased} and \ref{fig:TESS_GB_LCs_time} show the observed transit light curves. \autoref{tab:GB_obs_table} presents the ground-based observation log.  

\subsubsection{SPECULOOS-North and SPECULOOS-South observations}
We used the SPECULOOS \citep[Search for habitable Planets EClipsing ULtra-cOOl Stars,][]{Gillon2018, Delrez2018,Sebastian_2021AA,Burdanov2022} 1 meter network to observe five transits of TOI-7166.01. These observations were obtained with a 2K$\times$2K Andor iKon-L cameras with a pixel scale of $0.35\arcsec$ and a FOV of $12\arcmin\times12\arcmin$.
Three transits were observed with SPECULOOS-South located in Paranal, Chile.
The first transit was observed with SPECULOOS-South/Ganymede on UTC June 8, 2025 in the Sloan-$z'$ with an exposure time of 13\,s.
Second and third transits were observed on UTC June 20, 2025 with SPECULOOS-South/Io and SPECULOOS-South/Europa in the Sloan-$g'$ (exposure time of 140\,s) and Sloan-$z'$ (exposure time of 13\,s) filters, respectively.
SPECULOOS-North/Artemis observed two full transits in the Sloan-$g'$ and $I+z'$ filters on UTC July 3 and 29, 2025 with exposure time of 140\,s, and 13\,s, respectively.
Data processing and photometric measurements were performed using the {\tt PROSE}\footnote{{\tt Prose:} \url{https://github.com/lgrcia/prose}} pipeline \citep{prose_2022}.

\subsubsection{TRAPPIST-South observations}

We observed one full transit of TOI-7166.01 with the TRAPPIST-South \citep[TRAnsiting Planets and PlanetesImals Small Telescope,][]{Jehin2011,Gillon2011} telescope on UTC June 20, 2025. 
It is equipped with a 2K$\times$2K FLI Proline detector with a pixel scale of 0.65\arcsec and a FOV of $22\arcmin\times22\arcmin$.
The observations were conducted in the $R_c$ filter with and exposure time of 140\,s.

\subsubsection{LCOGT-1m0 observations}

The Las Cumbres Observatory Global Telescope (LCOGT; \citealt{Brown_2013}) 1.0-m network was used to observe three transits of TOI-7166.01.
The telescopes are equipped with 4096$\times$4096 SINISTRO Cameras, with an image scale of $0.389\arcsec$ per pixel and a FOV of $26^{\prime} \times 26^{\prime}$.
First transit was observed on UTC June 8, 2025 in the Sloan-$i'$, the second transit was observed on UTC July 3, 2025 in the Sloan-$r'$ (but the transit was not included in our global fit because of low S/R), and the last one was observed on UTC July 16, 2025 in the Sloan-$r'$ filter.
LCOGT data processing and photometric analysis were performed using {\tt BANZAI} pipeline \citep{McCully_2018SPIE10707E} and {\tt AstroImageJ} software \citep{Collins_2017}, respectively.

\subsubsection{TTT observations}
Three transits of TOI-7166.01 were observed with
the Two-meter Twin Telescope facility (TTT) during July 4, 16 and 29 nights. TTT is located at the Teide Observatory on the island of Tenerife (Canary Islands, Spain). Currently, it includes two 0.8m telescopes (TTT1 and TTT2) and a 2.0m telescope (TTT3) on altazimuth mounts. We used TTT1 telescope, that has two Nasmyth ports with focal ratios of $f/D=6.8$ and $f/D=4.4$ equipped with a QHY411M\footnote{\url{https://www.qhyccd.com/}} CMOS cameras \citep{2023PASP..135e5001A}. The QHY411M have  scientific Complementary Metal–Oxide–Semiconductor (sCMOS) image sensors with 14K x 10K 3.76~$\mu$m~pixel$^{-1}$ pixels. This setup, in the $f/D=6.8$ focus, provides an effective FoV of 30$^{\prime}\times$20$^{\prime}$ (with an angular resolution of 0.14"~pixel$^{-1}$). Science images were taken using the Sloan-$g'$ and  Sloan-$r'$ filters on UTC July 4, 2025.

TTT3 is a 2-m $f$/6 Ritchey-Chrétien telescope that is currently in its commissioning phase. An Andor iKon-L 936 2k$\times$2k camera is mounted at the Nasmyth 2 focus, equipped with a back-illuminated 13.5~$\mu$m pixel$^{-1}$ BEX2-DD CCD sensor, resulting in a field of view of 7.85$'$$\times$7.85$'$ and a plate scale of 0.23$''$~pixel$^{-1}$. Science images were taken using the Sloan-$g'$ and $y$   filters.
All the images were bias, dark and flat-field corrected in the standard way, and photometry extraction was performed using the {\tt PROSE} pipeline. The TTT1 and TTT2 data are not included in the final global analysis due to the low S/R and large photometric error bars.

\subsubsection{AUKR T80 observation}

We observed two transits of TOI-7166\,b on UTC July 29 and August 24, 2025 using the 80\,cm Prof. Dr. Berahitdin Albayrak Telescope  (T80) at the Ankara University Kreiken Observatory (AUKR), in the Sloan-$i'$ filter. The telescope is equipped with a $1024 \times 1024$ Apogee Alta U47+ CCD camera, providing a field of view of $11\arcmin \times 11\arcmin$. Data reduction and differential photometry were performed with the {\tt AstroImageJ} (AIJ) software \citep{Collins_2017}.

\subsubsection{OSN-1.5m observation}
We used the T150 at the Sierra Nevada Observatory in Granada (Spain) to observe one full transit of TOI-7166\,b on UTC 2025\,July\,29 in the Johnson-Cousin $I$ and $V$ filters. The telescope is equipped with a 2K$\times$2K Andor iKon-L BEX2DD CCD camera with a pixel scale of 0.232\arcsec, resulting in a total FOV of 7.9\arcmin$\times$7.9\arcmin. The data calibration and photometric extraction were performed using the {\tt PROSE} pipeline and the {\tt AstroImageJ} software.  Unfortunately, the data set was affected by adverse weather conditions and is therefore not included in the global analysis.

\subsection{High-resolution imaging} 
\label{obs:high-res}

To obtain high-resolution imaging for TOI-7166, we utilized the Zorro speckle interferometric instrument, mounted on the 8-m Gemini South telescope.
High-resolution imaging is critical to assess the local environment of an exoplanet host star and determine if a line of sight or bound close companion star is present. The presence of such a companion provides "third-light" contamination of the observed transit, leading to incorrect derived properties for the exoplanet and its host star \citep{Ciardi_2015,Furlan_2017AJ,Furlan_2020}.  

\begin{figure}
	\centering
     \includegraphics[scale=0.65]{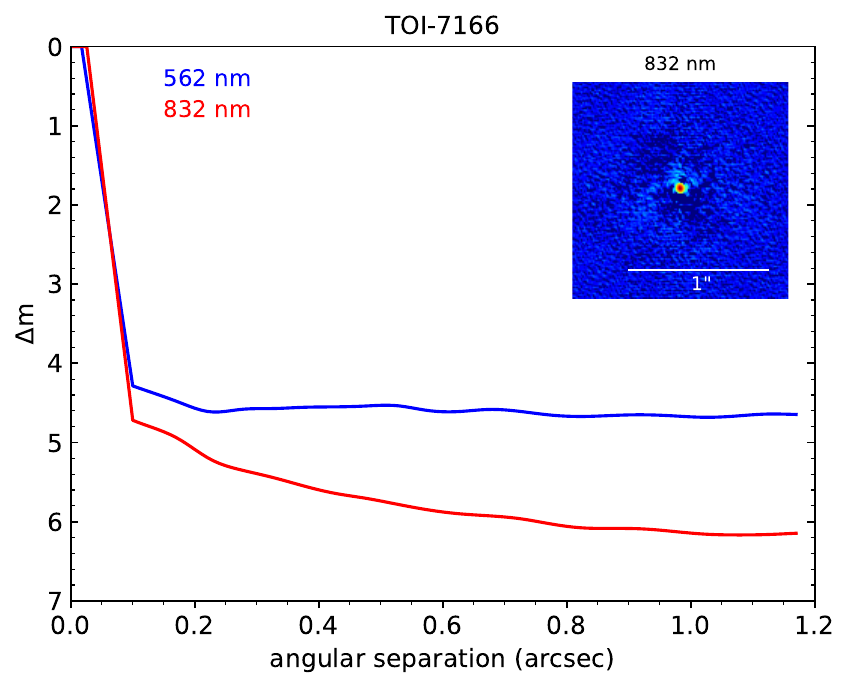} 
	\caption{The figure shows 5$\sigma$ magnitude contrast curves in both filters as a function of the angular separation out to 1.2 arcsec. The inset shows the reconstructed 832\,nm image of TOI-7166 with a 1 arcsec scale bar. TOI-7166 was found to have no close companions from the diffraction limit (0.02") out to 1.2 arcsec to within the magnitude contrast levels achieved.} 
	\label{fig:High_res_obs}
\end{figure}

TOI-7166 was observed with Zorro on UTC July 3, 2025.  Zorro provides simultaneous speckle imaging in two bands (562\,nm and 832\,nm), yielding output data products that include robust 5$\sigma$ magnitude contrast curves and a reconstructed image \citep{Scott_2021FrASS}.   Eight sets of 1000$\times$0.06 second images were obtained for TOI-7166 and the images were processed using our standard reduction pipeline \citep{Howell_2011AJ}. Figure~\ref{fig:High_res_obs} presents the final 5$\sigma$ magnitude contrast curves and the 832\,nm reconstructed speckle image for TOI-7166. We find that TOI-7166 is a single star with no companion brighter than 5-6 magnitudes below that of the target star from the Gemini Telescope 8-m telescope diffraction limit (20 mas) out to 1.2". At the distance of TOI-7166 (d=35.4 pc), these angular limits correspond to spatial limits of 0.7 to 42\,au.

\subsection{Spectroscopic data and stellar physical properties}
\label{sec:stellar_charac}

\begin{table}
\centering
	\caption{Astrometry, photometry, and spectroscopy stellar properties of TOI-7166. 
	(1): Gaia EDR3 \citealt{Gaia_Collaboration_2021A&A}; 
	(2) \emph{TESS} Input Catalog \citealt{Stassun_2018AJ_TESS_Catalog}; 
	(3) UCAC4 \citealt{Zacharias_2012yCat.1322};
	(4) 2MASS \citealt{Skrutskie_2006AJ_2MASS};
	(5) WISE \citealt{Cutri_2014yCat.2328}.
    }
	{\renewcommand{\arraystretch}{1.}
        \resizebox{0.5\textwidth}{!}{
		\begin{tabular}{lcc}
			\hline
			\hline
			\multicolumn{3}{c}{  Star information}   \\
			\hline
			{\it Target designations:}  & & \\
			            & \multicolumn{2}{l}{TOI 7166 }  \\
                        & \multicolumn{2}{l}{TIC 288421619 }  \\
			              & \multicolumn{2}{l}{GAIA DR3 1740534092250753920 } \\
			            & \multicolumn{2}{l}{2MASS J21224308+0853259}   \\
                        & \multicolumn{2}{l}{LP 577-37} \\
   			Parameter & Value &  Source   \\
			\hline
            \hline
			{\it Parallax  and distance:} &   \\
			RA [J2000]     &  21:22:43.35  &  (1) \\
			Dec [J2000]    & +08:53:21.83   &  (1)\\
			Plx [$mas$] & $28.382 \pm 0.022$ &   (1)\\
                $\mu_{RA}$ [mas yr$^{-1}$] & $260.21 \pm 0.02$ & (1) \\
                $\mu_{Dec}$ [mas yr$^{-1}$] & $-270.62 \pm 0.02$ & (1) \\
			Distance [pc]  & $35.23 \pm 0.03$ & (1)\\
			\hline
			{\it Photometric properties:} & \\
			TESS$_{\rm mag}$           &  $13.123 \pm 0.008$  & (2)  \\
			$V_{\rm mag}$ [UCAC4]       & $15.79 \pm 0.20$  & (3) \\
            $B_{\rm mag}$ [UCAC4]       & $16.8$  & (3) \\
            $R_{\rm mag}$ [UCAC4]       & $14.6$  & (3) \\
			$J_{\rm mag}$ [2MASS]       & $11.41 \pm 0.02$  &  (4) \\
			$H_{\rm mag}$ [2MASS]       & $10.86 \pm 0.02$  &    (4) \\
			$K_{\rm mag}$ [2MASS]       & $10.60 \pm 0.02$  &   (4) \\			
			$G_{\rm mag}$ [Gaia DR3]    & $14.492 \pm 0.001$  &  (1)  \\
			$W1_{\rm mag}$ [WISE]       & $10.39 \pm 0.02$ &  (5) \\
			$W2_{\rm mag}$ [WISE]       & $10.21 \pm 0.02$  &   (5) \\
			$W3_{\rm mag}$ [WISE]       & $10.05 \pm 0.06$  &   (5)\\
            $W4_{\rm mag}$ [WISE]       & $8.90$  &   (5)\\
			\hline
   			\multicolumn{2}{l}{\it Spectroscopic and derived parameters}  \\              
			  $T_{\rm eff}$ [K]              &  $  3099^{+51}_{-50} $  &    this work\\
			$\log g_\star$ [dex]           &  $  5.02 \pm 0.02 $  &  this work\\
			$\mathrm{[Fe/H]}$ [dex]       &  $  -0.20\pm 0.12 $  &   this work\\
			$M_\star$  [$M_\odot$]         & $  0.190 ^{+0.004}_{-0.004} $  &  this work\\
			$R_\star$  [$R_\odot$]         & $  0.222^{+0.006}_{-0.004} $ &   this work\\
            $L_\star$ [$L_\odot$]       & $0.004103 _{-0.000298}^{+0.000336}$  & this work  \\
			$F_{\rm bol}$  [erg s$^{-1}$ cm$^{-2}$]   &  $ ( 1.158 \pm 0.055) \times 10^{-10} $  &  this work\\
			$Av$ [mag]    &  $0.1 \pm 0.1$ & this work\\
			$\rho_\star$  [$\rho_\odot$]   & $ 17.36^{+0.96}_{-1.33}$  & this work \\
			$Age$  [Gyr]                   & $\lesssim$4~Gyr & this work \\
			Optical SpT                  & M4.5$\pm$0.5 & this work \\
			Near-infrared SpT            & M4$\pm$1 & this work \\
            \hline
	\end{tabular} }}
	\label{stellarpar}
\end{table}

\subsubsection{SED fitting}

\begin{figure}
	\centering
	\includegraphics[scale=0.3]{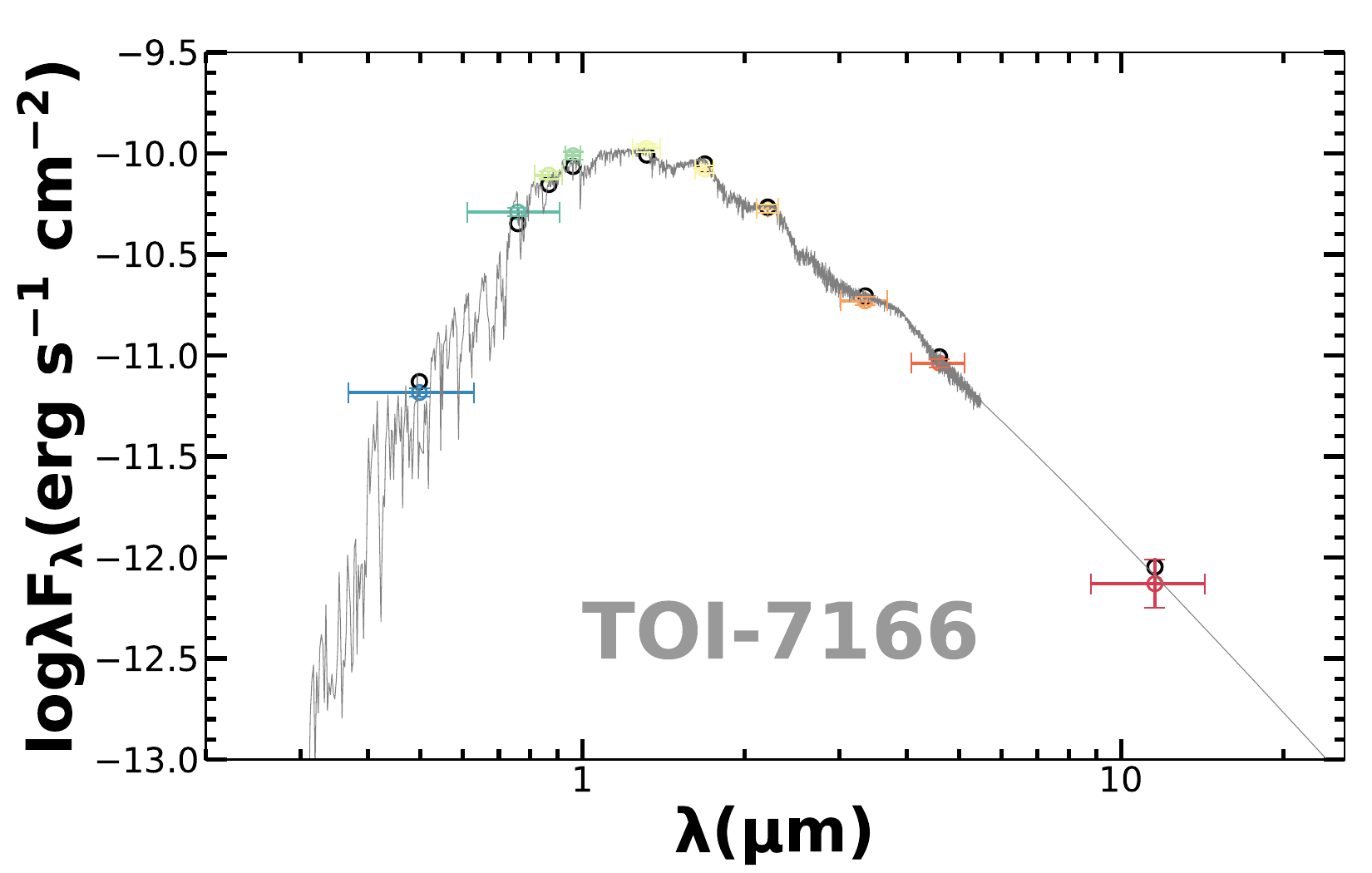}
	\caption{Spectral energy distribution of TOI-7166. The colored points with error bars represent the observed photometric measurements. The black circles are the model fluxes from the best-fit PHOENIX atmosphere model. The absolute flux-calibrated {\it Gaia} spectrophotometry is shown as the gray swathe.}
    	\label{fig:SED_plots}
\end{figure}

We performed an analysis of the broadband SED analysis of the star together with the {\it Gaia\/} DR3 parallax \citep[with no systematic offset applied; see, e.g.,][]{StassunTorres:2021}, in order to derive an empirical measurement of the stellar radius, following the same procedures described in \citet{Stassun:2016,Stassun:2017,Stassun:2018}. 
We pulled the near-infrared W1--W3 magnitudes from {\it WISE} together with the $zy$ magnitudes from {\it Pan-STARRS}, the $G_{\rm BP} G_{\rm RP}$ magnitudes from {\it Gaia}, and the $JHK_S$ magnitudes from {\it 2MASS}. 
We also utilized the absolute flux calibrated spectrophotometry from {\it Gaia}. Together, the available photometry spans the full stellar SED over the wavelength range 0.4--10~$\mu$m (see \autoref{fig:SED_plots}).  
 
We performed a fit using PHOENIX stellar atmosphere models \citep{Husser_2013}, with the free parameters being the effective temperature ($T_{\rm eff}$) and metallicity ([Fe/H]). The extinction, $A_V$, was fixed at zero due to the proximity of the system. The resulting fit (\autoref{fig:SED_plots}) has a reduced $\chi^2$ of 2.8, with a best-fit $T_{\rm eff} = 3100 \pm 75$~K, and [Fe/H] $= -0.1 \pm 0.2$. Integrating the model SED gives the bolometric flux at Earth, $F_{\rm bol} = 1.158 \pm 0.055 \times 10^{-10}$ erg~s$^{-1}$~cm$^{-2}$. Taking the $F_{\rm bol}$ together with the {\it Gaia\/} parallax directly gives the bolometric luminosity, $L_{\rm bol} = 0.00448 \pm 0.00021$~L$_\odot$. The stellar radius follows from the Stefan-Boltzmann relation, giving $R_\star = 0.232 \pm 0.013$~R$_\odot$. In addition, we can estimate the stellar mass from the empirical relations of \citet{Mann:2019}, giving $M_\star = 0.217 \pm 0.007$~M$_\odot$.

\subsubsection{IRTF/SpeX observations for TOI-7166}
\label{sec:irtf_spec}

\begin{figure}
    \centering
    \includegraphics[width=\linewidth]{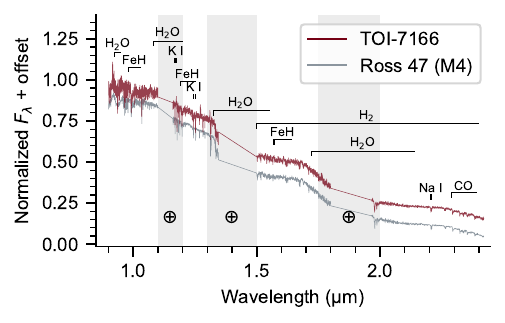}
    \caption{SpeX SXD spectrum of TOI-7166 (red) alongside the M4\,V standard Ross 47 (gray).
Prominent atomic and molecular features of M dwarfs are annotated, and regions of strong telluric absorption are shaded.
}
    \label{fig:spex}
\end{figure}

We observed TOI-7166 on 18 May 2025 with the SpeX spectrograph \citep{Rayner2003} on the 3.2-m NASA Infrared Telescope Facility (IRTF).
We used the short-wavelength cross-dispersed (SXD) mode and the $0\farcs3 \times 15''$ slit aligned to the parallactic angle to gather a spectrum covering 0.80–2.42\,$\mu$m with a resolving power of $R{\sim}2000$ and 2.5\,pixels per resolution element.
Conditions were clear with seeing of 0$\farcs$8’’.
We collected six 200\,s exposures at an airmass of 1.0, nodding in an ABBAAB pattern.
Science observations were preceded by six 20\,s exposures of the A0\,V telluric standard HD\,192538 (V=6.5) at a similar airmass and followed by the standard SXD calibration set.
Data reduction was carried out with Spextool v4.1 \citep{Cushing2004}, following the standard approach \citep[e.g.,][]{Ghachoui2023, Ghachoui2024, Barkaoui2023, Barkaoui2024, Barkaoui2025_TOI-6508b}.
The resulting spectrum (Fig.\,\ref{fig:spex}) has a median per-pixel S/R ratio of 79.

We analyzed the SpeX SXD spectrum of TOI-7166 using the SpeX Prism Library Analysis Toolkit \citep[SPLAT, ][]{splat} and referring to the IRTF Spectral Library \citep{Cushing2005, Rayner2009}.
The spectrum shows a strong match to the M4\,V standard Ross~47, and we adopt a near-infrared spectral type of M4.0 $\pm$ 1.0 accordingly.
Using the H2O--K2 index \citep{Rojas-Ayala2012} and $K$-band Na\,\textsc{i} and Ca\,\textsc{i} features in conjunction with the \citet{Mann2013} relation, we derive a stellar iron abundance of $\mathrm{[Fe/H]} = -0.20 \pm 0.12$, suggestive of subsolar metallicity.

\subsubsection{Shane/Kast observations for TOI-7166}
\label{sec:shane_kast}

\begin{figure}
    \centering
    \includegraphics[width=\linewidth]{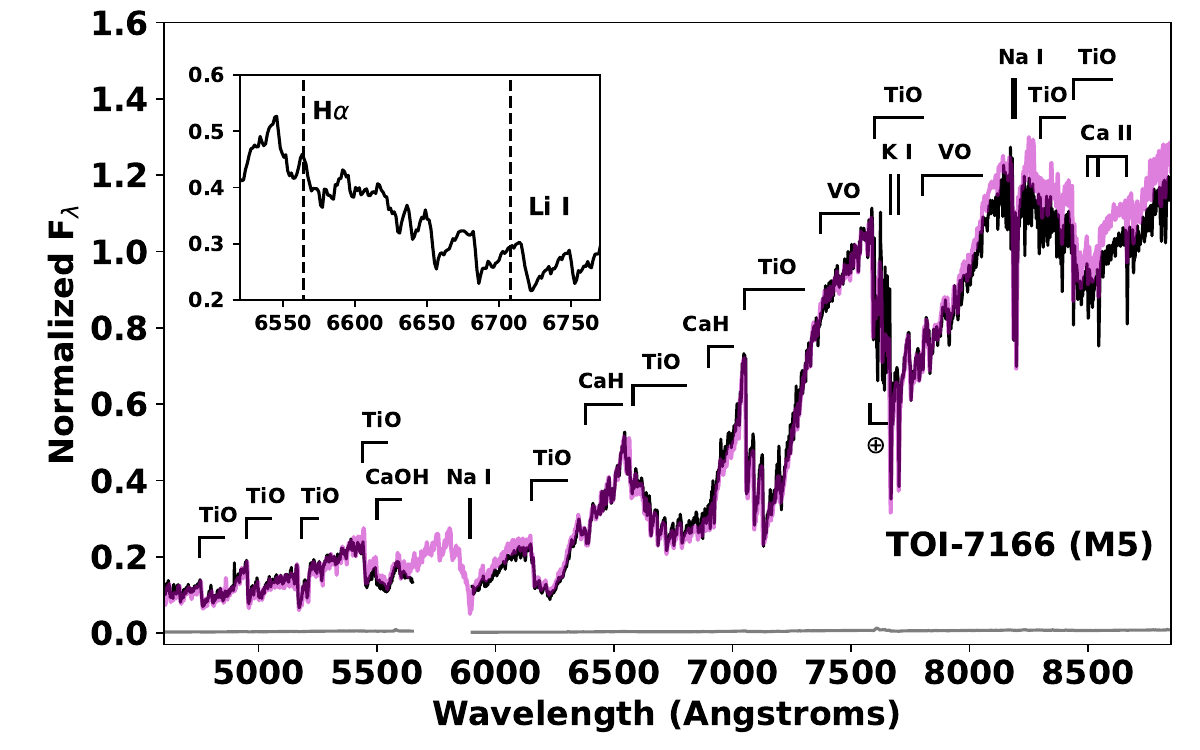}
\caption{Shane/Kast optical spectrum of TOI-7166 (black line) compared to its best-fit M5 SDSS spectral template from \citet[magenta line]{2007AJ....133..531B}. Key spectral features are labeled, including regions of residual telluric absorption ($\oplus$). 
The inset box shows the 6520--6770~{\AA} region encompassing H$\alpha$ (weak emission) and Li~I features (not present).
The gap in the Kast spectrum  between 5600~{\AA} and 5900~{\AA} corresponds to the gap between that instrument's blue and red channels.
}
    \label{fig:kast}
\end{figure}

We observed TOI-7166 with the Kast double spectrograph \citep{kastspectrograph} on the 3m Shane telescope at Lick Observatory on 27 July 2025 (UT) in clear conditions with 1$\farcs$2 seeing. We used the 1$\farcs$5 slit aligned to the parallactic angle to obtain blue and red optical spectra split at 5700~{\AA} by the d57 dichroic, and dispersed by the 600/4310 grism and 600/7500 grating, resulting in spectral resolutions of $\lambda/\Delta\lambda$ $\approx1100$ and $\approx1500$, respectively. We obtained a single 1200\,s exposure in the blue channel and two 600\,s exposures in the red channel at an average airmass of 1.2. The nearby G2~V star HD 211476 ($V=7.0$) was observed afterward at a similar airmass for telluric absorption calibration, and the spectrophotometric calibrator BD28~+4211 \citep{1990AJ.....99.1621O} was observed shortly thereafter for flux calibration. We used HeHgCd and HeNeArHg arc lamp exposures to wavelength calibrate our blue and red data, and flat-field lamp exposures for pixel response calibration. Data were reduced using the kastredux code\footnote{\url{https://github.com/aburgasser/kastredux}.} using standard settings. The resulting spectra have median signals-to-noise of
38 at 5425~{\AA} and 141 at 7350~{\AA}.

The reduced spectrum is shown in \autoref{fig:kast}, compared to the best-fit M5 dwarf SDSS spectral template from 
\citet{2007AJ....133..531B}.
TOI-7166 is slightly bluer than this template, indicating a slightly earlier type.
Index-based classifications based on methods described in
\citet{1995AJ....110.1838R,1997AJ....113..806G,1999AJ....118.2466M,2003AJ....125.1598L}; and \citet{2007MNRAS.381.1067R} indicate M4.5$\pm$0.5 as a more accurate spectral type that encompasses the near-infrared classification.
We detect weak H$\beta$ (4861~{\AA}) and H$\alpha$ (6563~{\AA}) emission, the latter with an equivalent width $\mathrm{EW} = -1.21\pm0.12$~{\AA}, corresponding to $\log{\left(L_{{\rm H}\alpha}/L_{\rm bol}\right)} = -4.52\pm0.08$ using the $\chi$ factor relation of \citet{2014ApJ...795..161D}. 
The presence of weak H$\alpha$ emission indicates an activity age $\lesssim$7~Gyr based on the kinematic sample of \citet{2008AJ....135..785W}, or $\lesssim$4~Gyr based on the mass-dependent relation of \citet{2024ApJ...966..231P} assuming M = 0.15~M$_\odot$. The absence of detectable Li\,\textsc{i} absorption at 6708~{\AA} rules out a substellar mass and age less than $\sim30$\,Myr. We measure the metallicity index $\zeta=1.080\pm0.005$ \citep{2013AJ....145..102L}, which corresponds to a roughly solar metallicity of [Fe/H]$=+0.11\pm0.20$ using the \citet{Mann2013} calibration, albeit formally consistent with the slightly subsolar metallicity inferred from the near-infrared spectrum. \\

\section{Planet validation}
\label{sec:planet_validation}
\subsection{TESS data report}
The transit search was performed with the SPOC pipeline \citep{jenkins2002,jenkins2010,Jenkins2020} using Sector 82 data on September 13, 2024. The pipeline found a transit signal at 12.92 days with a signal-to-noise ratio of 10.5. The TESS Science Office alerted the event on November 14, 2024 \citep{guerrero_TOIs2021ApJS}. The transit signal passed all diagnostic tests presented in the data validation reports \citep{Twicken_DVdiagnostics2018}. The source of the transits was localized to $2.08\pm2.97$\arcsec from TOI-7166.
It resulted in a transit depth of $8723.2564 \pm 870.9261$~ppm, duration of $1.5980 \pm 0.2144$~hrs, and an orbital period of $12.9228 \pm 0.0030$~days, which correspond to a planet with a radius of $R_p = 2.1 \pm 0.32$~$R_\oplus$.

\subsection{Ground-based photometric follow-up for TOI-7166}
\label{sec:GB_photo_foll}
\begin{figure}
	\centering
	\includegraphics[scale=0.2]{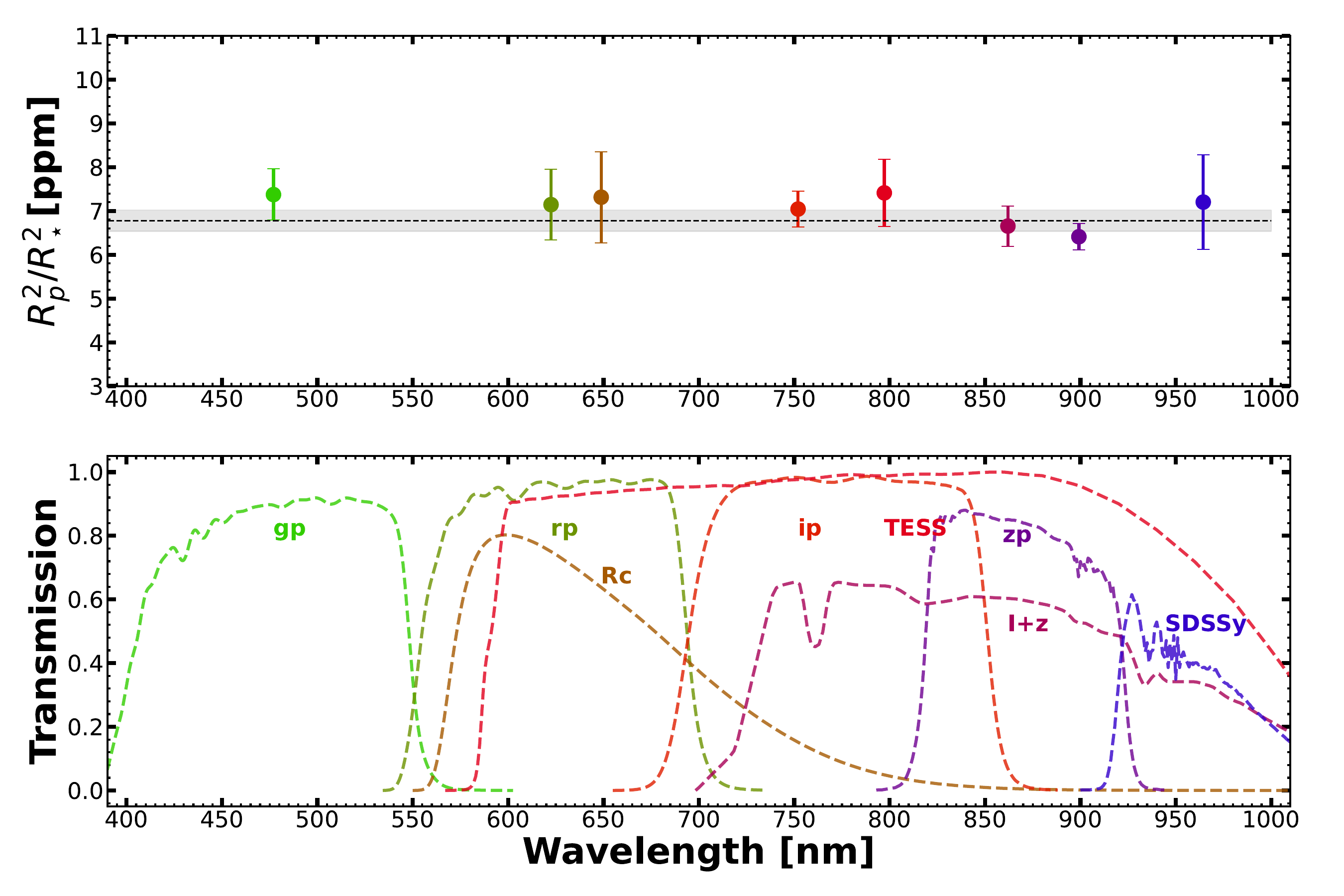}
	\caption{{\it Top panel:} Measured transit depths of the planet in different filters (colored dots highlighted with error bars) obtained from our global analysis for  TOI-7166\,b. The horizontal black  line  corresponds to the depth obtained from the achromatic fit with a $1\sigma$ error bar (shaded region). All measurements agree with the common transit depth at $1\sigma$. {\it Bottom panel:} Transmission for each filter.
    }
	\label{toi2015b_depths}
\end{figure}

We used the ground-based photometric observations to {\it i}) confirm the event on the target, {\it ii)} refine the transit ephemerides and {\it iii}) measure the transit depth in different bands in order to validate the planetary nature. 
We conducted time-series observations in different bands, Sloan-$g'$, -$r'$, -$i'$, -$z'$, SDSS$y$, and $Rc$ filters, covering a wavelength range from 4000 to 10000\,\AA. The aperture photometry was performed in uncontaminated small apertures of only a few arc-seconds to exclude any neighborhood objects. It resulted in no chromatic dependence across filters. Figure~\ref{toi2015b_depths} shows the measured transit depths in different bands.

\subsection{Archival data for TOI-7166}

\begin{figure*}
	\centering
	\includegraphics[scale=0.7]{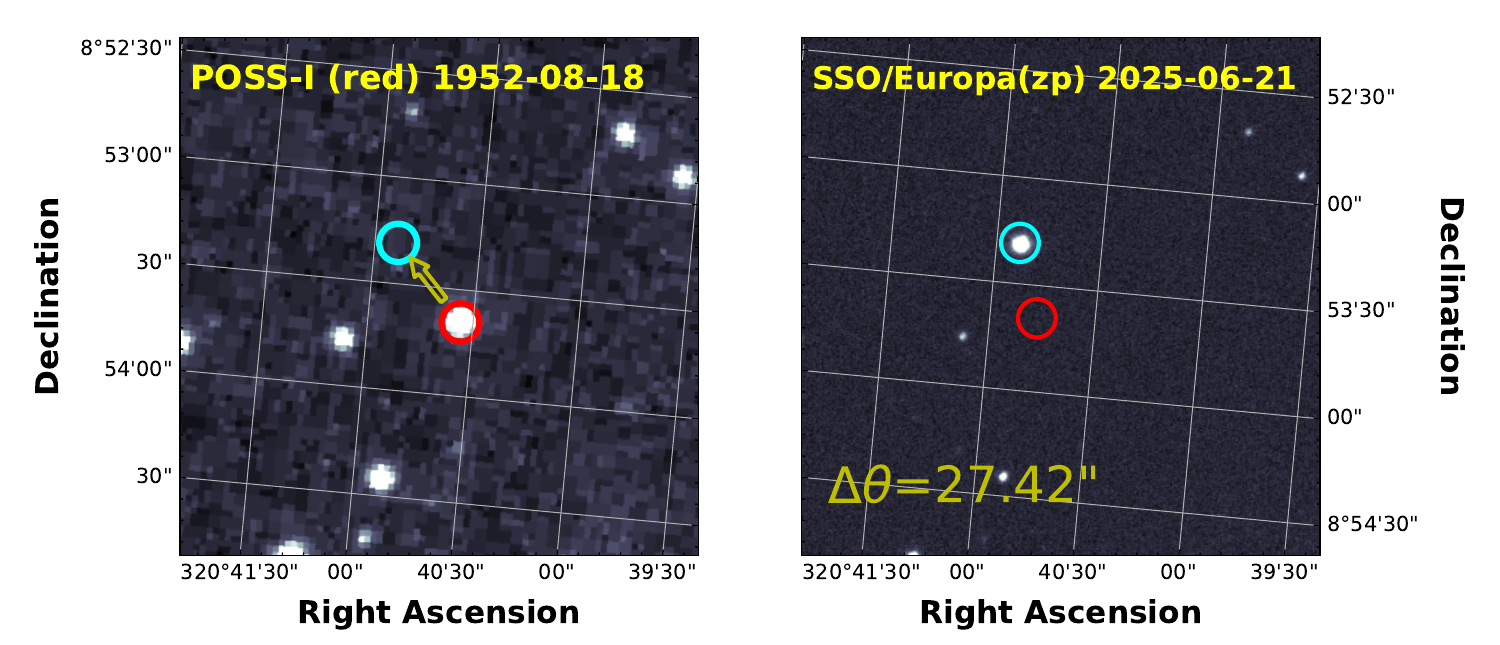}
	\caption{Evolution of  TOI-7155 position.  The {\it left panel} shows the red image from POSS-I taken in 1952. The {\it right panel} shows the \textit{zp} image from SPECULOOS-South/Europa taken in 2025. The previous and current positions of the target are shown in red and blue circles, respectively.}
	\label{archival_images}
\end{figure*}

We explored the archival science data for TOI-7166 to exclude any possible background stellar objects that could be blended with TOI-7166 in its current position. 
TOI-7166 has a relative high proper motion of 375.7\,mas/yr. We used the data from POSS-I/DSS \citep{1963POSS-I} in 1952 in the red filter and SPECULOOS-South in 2025 in the Sloan-$z'$ filter, and spanning 73 years. The target has shifted by 27\farcs{4} from 1952 to 2025. Fortunately, no background objects are detected in the current position of TOI-7166 (see Figure~\ref{archival_images}). 

\subsection{Statistical validation of TOI-7166.01}

We used the {\tt TRICERATOPS}\footnote{{\tt TRICERATOPS:}~\url{https://github.com/stevengiacalone/triceratops}} \citep{Giacalone_2021AJ} package developed in Python to compute the False Positive Probability, which allows us to identify whether a given candidate is a planet or a nearby false positive. {\tt TRICERATOPS} provides two output parameters, which are the FPP (False Positive Probability) and the NFPP (Nearby False Positive Probability).
{\tt TRICERATOPS} uses the phase-folded \emph{TESS} or ground-based light curves of the candidate together with high-contrast imaging observations in order to improve our results.
In this case, we used photometric observation from TESS sector 82 (Section~\ref{sec:tess_photometry}), and high-resolution observation from Gemini-South/Zorro taken on UTC July 3, 2025 (Section~\ref{obs:high-res}).
We obtained NFPP$=0$ (i.e., the event was detected on the target, Section~\ref{sec:GB_photo_foll}) and FPP$= 0.0018 \pm 0.0005$. TOI-7166\,b is validated as a planet.

\section{Global analysis of photometric data}
\label{sec:glbal_fit}


\begin{table}
\caption{Derived properties of the TOI-7166\,b system with 1-$\sigma$.}
	\begin{center}
		{\renewcommand{\arraystretch}{0.9}
				\resizebox{0.48\textwidth}{!}{
			\begin{tabular}{ll}
		\hline
            \multicolumn{2}{c}{TOI-7166}  \\
        \hline
				Parameter &  Value    \\   
	    \hline
            \multicolumn{2}{l}{\it Quadratic Limb-Darkening coefficients}  \\
        \hline
			    $u_{\rm 1,TESS}$ &  $0.31 \pm 0.01$   \\
				$u_{\rm 2,TESS}$ &  $0.23 \pm 0.05$   \\
				$u_{\rm 1,{\rm Sloan-z'}}$ &  $0.24 \pm 0.02$   \\
				$u_{\rm 2,{\rm Sloan-z'}}$ &  $0.18 \pm 0.05$  \\
                $u_{\rm 1,{\rm I+z'}}$ &  $0.26 \pm 0.02$   \\
				$u_{\rm 2,{\rm I+z'}}$ &  $0.20 \pm 0.04$  \\
                $u_{\rm 1,{\rm SDSSy}}$ &  $0.22 \pm 0.01$   \\
				$u_{\rm 2,{\rm SDSSy}}$ &  $0.17 \pm 0.04$  \\
				$u_{\rm 1,{\rm Sloan}-i'}$ &  $0.34 \pm 0.02$   \\
				$u_{\rm 2,{\rm Sloan}-i'}$ &  $0.23 \pm 0.06$   \\
				$u_{\rm 1,{\rm Sloan}-r'}$ &  $0.54 \pm 0.04$  \\
				$u_{\rm 2,{\rm Sloan}-r'}$ &  $0.26 \pm 0.06$   \\
                $u_{\rm 1,{\rm Sloan}-g'}$ &  $0.57 \pm 0.03$  \\
				$u_{\rm 2,{\rm Sloan}-g'}$ &  $0.34 \pm 0.06$   \\
                $u_{\rm 1,{\rm Johnson}-Rc}$ &  $0.38 \pm 0.02$  \\
				$u_{\rm 2,{\rm Johnson}-Rc}$ &  $0.26 \pm 0.04$   \\
        \hline
        \multicolumn{2}{l}{\it Derived planet parameters}  \\
        \hline
            Orbital period $P$ [days]                &  $12.920636210 ^{+0.000000998}_{-0.000000998}$   \\ 
            Transit depth $R^2_p/R^2_\star$ [ppt]    &  $ 6776^{+198}_{-194} $   \\
            planet-to-star ratio $R_p/R_\star$       & $0.0823 \pm 0.0012$ \\
            Planet radius $R_p$ [$R_\oplus$]         &  $2.01^{+0.06}_{-0.05}$      \\  
            Transit-timing $T_0$                     &  $ 10847.7427718 \pm 0.0000101$     \\
            $[{\rm BJD}_{\rm TDB} - 2450000]$        &     \\
            Scaled semi-major axis  $a/R_\star$      &  $ 59.98^{+1.09}_{-1.57}$     \\
            Orbital semi-major axis $a$ [AU]         &  $ 0.06191 \pm 0.00044$    \\
            Orbital inclination $i$ [deg]            &  $ 89.80^{+1.16}_{-1.05}$    \\
            Impact parameter $b$ [$R_\star$]         &  $0.20^{+0.10}_{-0.12} $    \\
			Transit duration  [min]               &  $ 105 \pm 1 $    \\
            Eccentricity $e$                         &  $ 0$ [fixed] \\
            Equilibrium temperature $T_{\rm eq}$ [K] &  $249 \pm 5$      \\ 
            Incident flux $<F>$  [$<F_\oplus>$]                    & $ 1.07 ^{+0.09}_{-0.08} $ \\
   \hline
		\end{tabular}}}
	\end{center}
	\label{tab:mcmc_results}
\end{table}

Our global modeling of transit observations is based on the TESS (described Section~\ref{sec:tess_photometry}) and ground-based data (described in Section~\ref{sec:gb_photometry}), using the Metropolis-Hastings (MH) \citep{Metropolis_1953,Hastings_1970} technique implemented in {\tt Trafit}, a revised version of the Markov chain Monte Carlo (MCMC) code (see \cite{Gillon2014AA} for more details, and references therein). We followed the same strategy as described in \citet{Barkaoui2023,Barkaoui2024,Barkaoui2025_TOI-6508b}.

The transit data  are fitted using the \cite{Mandel2002} quadratic limb-darkening model, multiplied by a transit baseline, to correct systematic effects (time, FWHM, Airmass and background). For each transit light curve, the baseline is selected by minimizing the Bayesian information criterion (BIC; \citet{schwarz1978}). The photometric measurement error bars are re-scaled using the correction factor $CF = \beta_{w} \times \beta_{r}$, where $\beta_{r}$ is the red noise and $\beta_{w}$ is the white noise \citep{Gillon2012}. 

For the global fit, the free parameters for transit modeling used are the orbital period, total transit duration, impact parameters, transit depth, and stellar density.
We applied a Gaussian prior distribution to the stellar effective temperature ($T_{\rm eff}$), surface gravity ($\log g_\star$), mass ($M_\star$), radius ($R_\star$), metallicity ($\mathrm{[Fe/H]}$) and quadratic limb-darkening coefficients $u_1$ and $u_2$ (see Table~\ref{tab:priors}). Given $T_{\rm eff}$, $\mathrm{[Fe/H]}$ and $\log g_\star$, we computed the coefficients $u_1$ and $u_2$ using {\tt LDTk} \footnote{{\tt LDTk:} \url{https://github.com/hpparvi/ldtk}} package \citep{Parviainen2015}.
During the fitting, we converted $u_1$ and $u_2$ coefficients into $q_1 = (u_1 + u_2)^2$ and $q_2 = 0.5u_1(u_1 + u_2)^{-1}$ proposed by \citealt{Kipping_2013MNRAS.435.2152K}.

Two global MCMC analysis were performed. First one assuming a circular ($e=0$) orbit and second one assuming an eccentric orbit. Our results favored a circular orbit solution based on the Bayes factor.
For each transit light curve, we performed a  preliminary fit composed of one Markov chain with $5\times 10^5$ steps to determine the correction factor ($CF$; \citet{Gillon2012}) to be applied to the measurements' error bars. Then, we performed a final MCMC fit  composed of five Markov chains with one million steps to constrain the final physical parameters of the system. We used the \cite{Gelman1992} statistical tests to check the convergence for each Markov chain.
Our final solution for the circular orbit is presented in Table~\ref{tab:mcmc_results}.

\section{Planet searches and detection limits from the TESS photometry}
\label{sec:planet_searches}

Using the available TESS data (see Section~\ref{sec:tess_photometry}), we employed the \texttt{SHERLOCK} package \citep{pozuelos2020,demory2020}, to independently recover the candidate TOI-7166.01 and to explore the existence of additional signals that may have been missed by the official SPOC and QLP pipelines \citep[see, e.g.,][]{geo2024,seba2025,seluck2025}. \texttt{SHERLOCK} is specifically designed to identify low-S/N transit-like features potentially attributable to planets, and it provides tools for candidate validation and preliminary characterization, as described in \citet{devora2024}.

We first recovered the signal corresponding to the planetary candidate alerted by SPOC, TOI-7166.01, and a secondary signal that would correspond to an orbital period of 7.03\,d and $\sim$1.20\,R$_{\oplus}$. We executed the vetting module for this secondary candidate and found no obvious false-positive source that may have produced the signal. In addition, \texttt{SHERLOCK} relies on TRICERATOPS \citep{Giacalone_2021AJ} to conduct a statistical validation; in this case, the FPP and NFPP values were 0.42 and 0.09, respectively, placing the candidate in the ambiguous area out of the likely planet region, and close to the border with the nearby false positive area. Hence, to finally validate or refute this signal, we triggered a ground-based campaign using SPECULOOS-North/Artemis, LCO-SSO-2m0/MuSCAT4, and LCOGT-1m0 telescopes, which resulted in no detection in any of our four trials, suggesting this signal is a false positive. 

The lack of additional candidates may be explained by several possibilities \citep[see, e.g.,][]{wells2021,Schanche2021,pozuelos2023}: (1) the system hosts no further planets; (2) additional planets are present but do not transit; (3) additional transiting planets exist, yet their orbital periods exceed the range investigated in this study; or (4) further transiting planets are present, but their signals remain undetectable due to the limited photometric precision of the available data. Scenarios (1) and (2) could be addressed with a high-precision radial velocity campaign, although such an analysis lies beyond the scope of this work. Scenario (3) can be tested by extending the observational baseline; unfortunately, no other TESS observations are planned. To explore the fourth scenario, we conducted injection-and-recovery experiments with the \texttt{MATRIX} code\footnote{{The \texttt{MATRIX} (\textbf{M}ulti-ph\textbf{A}se \textbf{T}ransits \textbf{R}ecovery from \textbf{I}njected e\textbf{X}oplanets) package is open access on GitHub: \url{https://github.com/PlanetHunters/tkmatrix}}} \citep{devora2022}.  

\texttt{MATRIX} explores a three-dimensional parameter space by building a grid of orbital periods, planetary radii, and transit epochs. Each 3-parameter set defines a synthetic transit signal that is injected into the original TESS light curve. In our analysis, we adopted a grid of 60 orbital periods (1–15\,d), 60 planetary radii (0.5–3\,$R_{\oplus}$), and 5 transit epochs, yielding a total of 18,000 scenarios. For each of these scenarios, we applied a detrend with a bi-weight filter using a window size of 0.5\,d, which was found to be the optimal length during the \texttt{SHERLOCK} exploration. Then, the light curves are processed in the search for planets, where a synthetic planet is considered as retrieved when its found period and epoch differ by at most 1\% and up to 1 hour from the injected values, respectively.

\begin{figure}
    \centering
    \includegraphics[width=\columnwidth]{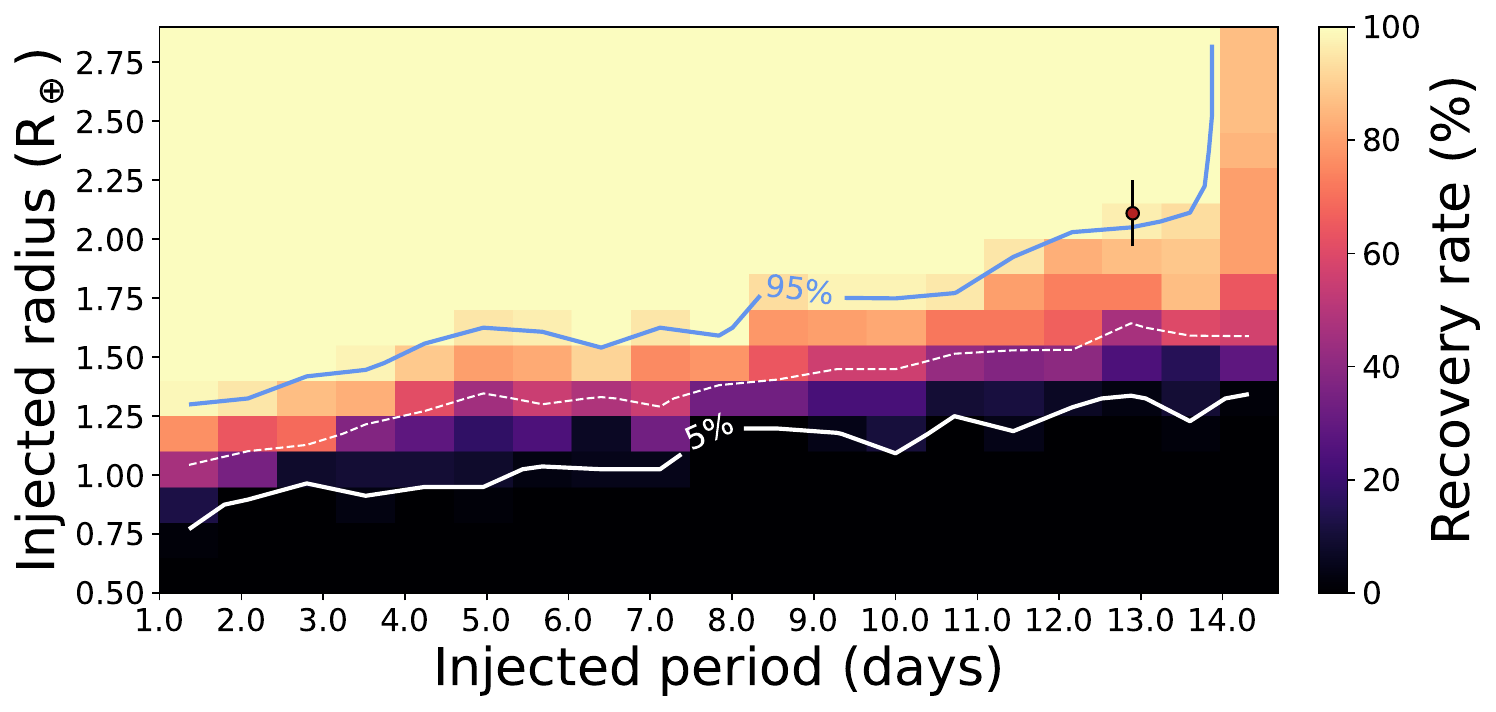}
    \caption{Results of the injection-recovery experiment conducted with MATRIX to determine the detectability of the planets in the TESS data. The colour scale represents recovery rates, where bright yellow indicates high recovery and dark purple/black indicates low recovery. The solid blue line marks the 95\% recovery contour, the dashed white line indicates the 50\%, and the solid white line shows the 5\%. The red dot marks the nominal value for TOI-7166\,b.}
    \label{inj-rec}
\end{figure}

The results are displayed in Figure~\ref{inj-rec}. The transition region, that is, the region with recoveries of 50\%, gently increases from 1\,R$_{\oplus}$ at 1\,d orbital period to 1.5\,R$_{\oplus}$ at 15\,d orbital period. We found that Earth-sized planets become invisible for orbital periods longer than 3\,d, and planets larger than 1.5\,R$_{\oplus}$ would be easily detectable for the full range of periods studied here, with recovery rates of 80--100\%; hence confirming the non-existence of any. 

Additionally, we computed the Lomb--Scargle periodogram \citep{Lomb_1976ApSS,Scargle_1982ApJ}, which showed no indications of flaring activity or stellar rotation modulation in the TESS Sector 82 data of the target. This implies that the rotational period of the host star is probably longer than the TESS observation window for a single sector.

\section{Prospects for further follow-ups}
\label{sec:prospects}

\subsection{Planetary mass determination}
\label{sec:mass_determ}

Recent studies have quantified the observational thresholds required to infer planetary interiors from mass and radius measurements \citep[see, e.g.,][]{dorn2015,dorn2017,Plotnykov2020}. From our best fit model, the measured radius of TOI-7166\,b is $2.01^{+0.06}_{-0.05}$\,R$_\oplus$ (relative precision $\sim$2.7\%). Following \citet{plotnykov2024}, assuming a rocky Earth-like planet, we would need a mass precision of 7–17\% for achieving iron-mass fractions and core-mass fractions within $\pm$10–15\,wt\%. In the case of assuming a water-rich planet, the mass precision would be 5--15\% to keep the water-mass fraction within $\pm$10–15\,wt\%. Then, we can adopt a conservative mass precision target of 15\% as the threshold required to enable a reliable first-order characterization of the planet’s internal structure.

Hence, to quantify the radial-velocity follow-up efforts required to measure the mass of TOI-7166\,b with such a precision, we conducted a dedicated suite of simulations. To this end, we generated synthetic RV time series by randomly sampling observation dates within the allowed visibility window and evaluating the Keplerian signal of the known planet assuming a planetary mass of $5.03^{+1.54}_{-1.04}$\,M$_{\oplus}$ derived by the \spright code \citep{spright2024}. In the absence of prior RVs for this target, the per-epoch uncertainty was modeled using an effective error ($\sigma_{eff}$) adopted directly from the literature for stellar analogs (similar spectral type, brightness, and T{$_{eff}$) observed considering two state-of-the-art facilities: CARMENES and MAROON-X \citep[see, e.g,][]{reiners2018,Barkaoui2025_TOI2015bc}. 

We use N to represent the number of RV measurements in a planned observing campaign. For each set of N observations, we calculated the noiseless Keplerian RVs and then created 100 Monte Carlo simulations by adding Gaussian noise with a standard deviation of $\sigma_{eff}$. Then, each set of N simulated observations was fitted using the \curvefit function from the scipy library \citep{2020SciPy-NMeth}, treating planetary mass as a free parameter. While this approach is simpler than a full multi-parameter Bayesian analysis, it enables us to roughly estimate how mass uncertainty changes with the number of measurements and helps approximate the number of observations required to achieve a desired mass precision. For simplicity, this method does not account for correlated noise, stellar activity, or multiple-parameter relationships. As a consequence, the derived number of measurements should be regarded as a lower limit, since the absence of these noise sources makes the simulation more optimistic than what is typically achievable in real observations. Nevertheless, during the calibration of this methodology by comparing simulated and real data, we found that for quiet stars, i.e., those with low levels of stellar activity (such as TOI-7166), the deviation between our model predictions and the actual mass determination is typically $<$3\%. This provides confidence in the robustness and reliability of the procedure.

Using this method, we find that CARMENES would require approximately 400 observations to achieve the 15\% precision goal, which aligns with expectations given the target’s faintness for this instrument \citep[see, e.g.,][]{ribas2023}. Since the object is visible for about five months each year, this would mean an observing campaign lasting around three observational semesters. In comparison, MAROON-X can achieve the same precision with only about 15 to 20 measurements, which could be completed in a single semester. This makes MAROON-X the optimal instrument for mass determination and enabling interior modeling for TOI-7166\,b.

\subsection{Atmospheric characterization}

To quantify the suitability of transiting exoplanets for atmospheric characterization using the transmission spectroscopic observation, we used the transmission spectroscopy metric (TSM) introduced by \cite{kem}. By combining the planetary parameters (radius $R_p$, mass $M_p$ estimated from \cite{spright2024}, and equilibrium temperature $T_{\rm eq}$), together with the infrared brightness of the host star $J_{\rm mag}$, we find that TOI-7166\,b has a TSM of $62^{+21}_{-14}$.
Right panel of Figure\,\ref{fig:Teff_Sp_TSM_Teq} shows the TSM against the planetary equilibrium temperature for known transiting exoplanets with mass measurements and radius $R_p<4R_\oplus$ and stellar effective temperature $T_{\rm eff}<4000$K. The diagram shows that TOI-7166\,b is a suitable sub-Neptune-sized planet for detailed atmospheric characterization with the \emph{JWST}.

\section{Conclusion}
\label{sec:conclusion}

We present the validation and discovery of TOI-7166\,b system by the TESS mission. The system has been confirmed using multi-band photometric observations collected with the SPECULOOS-North, SPECULOOS-South-1m0, TTT-2m0, TRAPPIST-South-0.6m and LCOGT-1m0 telescopes (see \autoref{sec:gb_photometry}).
We characterized the target by combining the spectral energy distribution (SED) together with
spectroscopic observations obtained with IRTF/SpeX (Section~\ref{sec:irtf_spec}) Shane/Kast (Section~\ref{sec:shane_kast}) instruments.
We performed a global fit of the TESS data and ground-based multi-color observations to derive and constrain the physical properties of the TOI-7166 system (see Section~\ref{sec:glbal_fit}).
\autoref{stellarpar} shows the stellar properties (astrometric, photometric and spectroscopic) of the target star.
\autoref{tab:mcmc_results} shows the derived physical parameters of the system. The  posterior distribution parameters of the system are shown in \autoref{fig:Corner_plot}.

We find that TOI-7166 is a nearby M4-type at a distance of $d=35.2$~pc, with an effective temperature of $T_{\rm eff} = 3099 \pm 50$K, a stellar mass of $M_\star = 0.190 \pm 0.004  M_\odot$, and a stellar radius of $R_\star = 0.222 \pm 0.005R_\odot$, a surface gravity of $\log g_\star = 5.02\pm 0.02$~dex and a metallicity of $\mathrm{[Fe/H]} = -0.20 \pm 0.12$~dex.
TOI-7166\,b is a mini-Neptune-sized planet completes its orbit in 12.92~days which places it close to the inner edge of the Habitable zone of its host star. It has \ a planetary radius of $R_p = 2.01^{+0.06}_{-0.05}$R$_\oplus$, an equilibrium temperature of $T_{\rm eq} = 249 \pm 5$K (assuming a null Bond Albedo), and an insolation of $S_p = 1.07 \pm 0.08$S$_\oplus$.

The predicted radial velocity amplitude using the \cite{spright2024}'s  mass-radius relationship is found to be $K_{\rm RV} = 4.24^{+2.11}_{-1.12}$m/s. By combining the brightness of the star ($V_{\rm mag}  = 15.8$) and the predicted radial velocity amplitude, which make TOI-7166 a suitable target for radial velocity spectroscopic observation follow-up using the MAROON-X spectrograph (Section~\ref{sec:mass_determ}). We have observed a similar target with MAROON-X, TOI-2015 ($V_{\rm mag} = 16.1$). Radial velocity measurements are presented in \cite{Barkaoui2025_TOI2015bc}.
Moreover,  combining the infrared brightness of the star ($J_{\rm mag} = 11.4$ and $K_{\rm mag} = 10.6$) together with the planet-to-star ratio $R_p/R_\star = 0.0823 \pm 0.0012$ makes TOI-7166\,b a favorable target for upcoming \emph{JWST} observations for transmission spectroscopy.

\begin{figure*}
    \centering
    \includegraphics[width=\columnwidth]{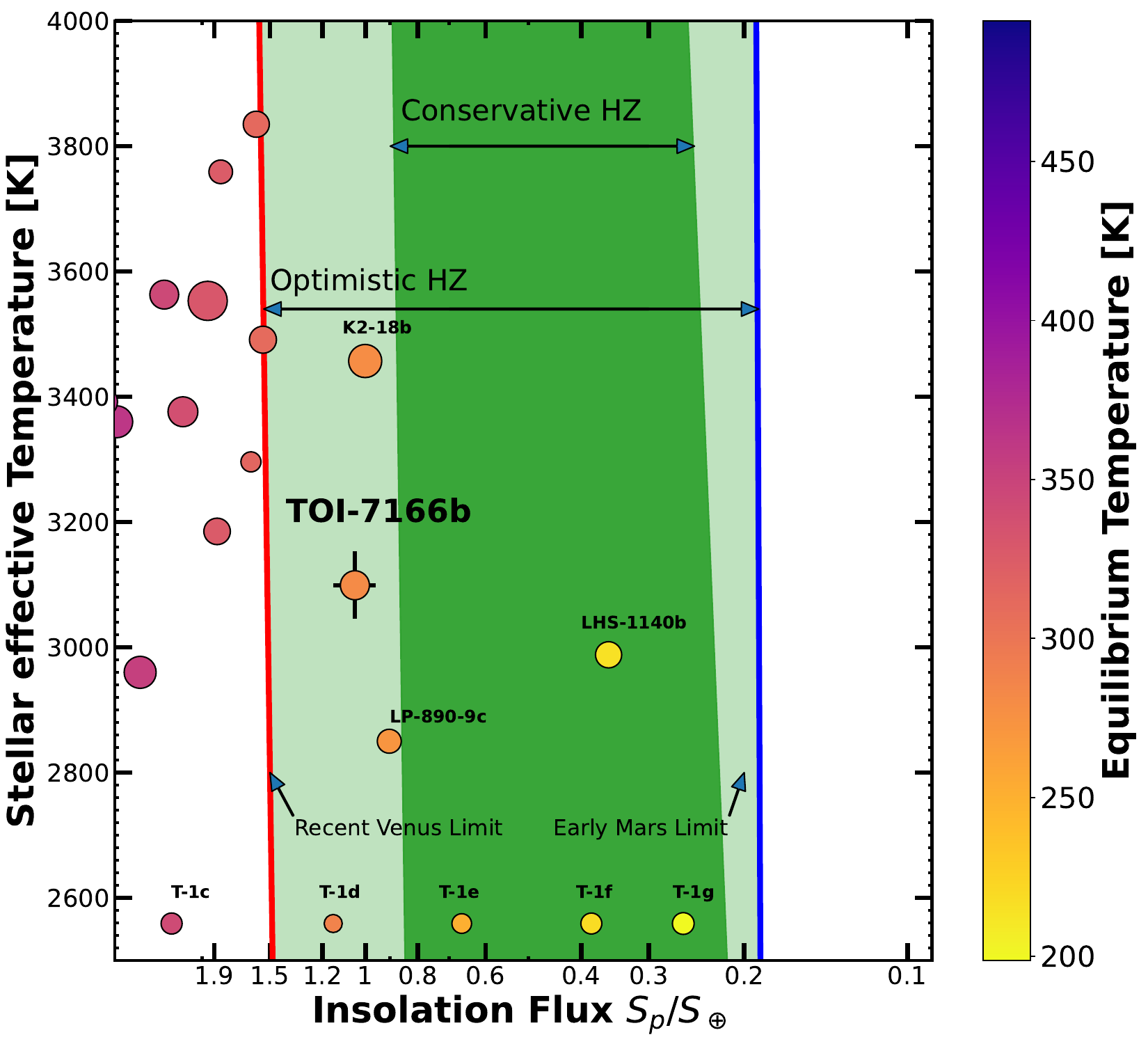}
    \includegraphics[width=\columnwidth]{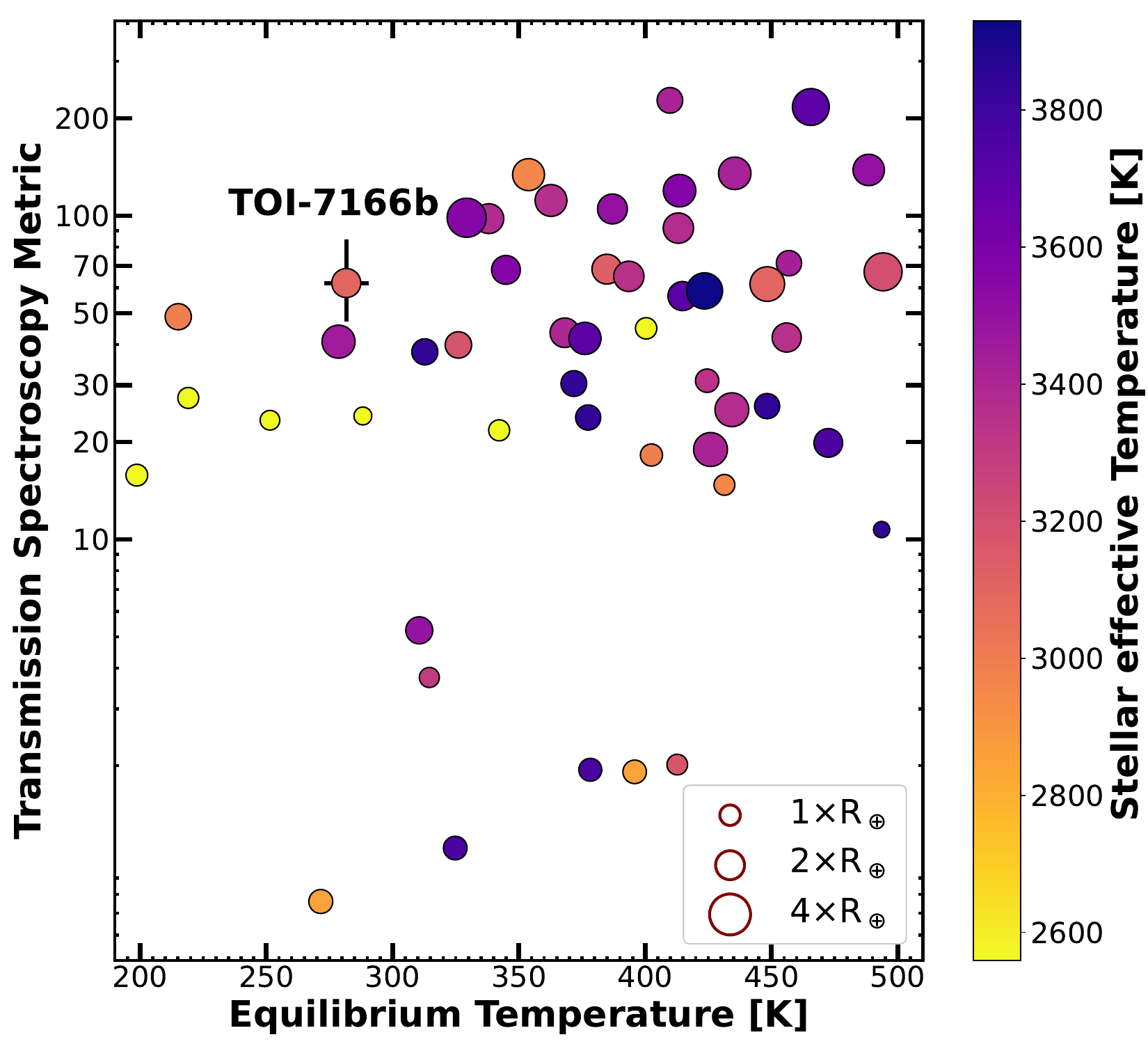}
    \caption{{\it Left panel:} Stellar effective temperature ($T_{\rm eff}$) as a function of incident stellar flux ($S_p$) of known transiting exoplanets orbiting host stars cooler than $4000$\,K. The size of each point corresponds to the planet's size, and the color indicates its equilibrium temperature. The light green region denotes the optimistic habitable zone, bounded by a solid red line (recent Venus limit) and a solid blue line (Early Mars limit). The dark green region indicates the conservative habitable zone as defined by \citealt{Kopparapu_2013ApJ}. 
    \textit{Right panel:} Transmission spectroscopy metric \citep{kem} against the planetary equilibrium temperature for the same sample displayed in the left panel. The points are colored according to the stellar effective temperature. TOI-7166\,b is highlighted by the error bars.
    }
    \label{fig:Teff_Sp_TSM_Teq}
\end{figure*}

\section*{Acknowledgments}
Funding for KB was provided by the European Union (ERC AdG SUBSTELLAR, GA 101054354).
Author F.J.P acknowledges financial support from the Severo Ochoa grant CEX2021-001131-S funded by MCIN/AEI/10.13039/501100011033 and 
Ministerio de Ciencia e Innovación through the project PID2022-137241NB-C43. Partially based on observations made at the Observatorio de Sierra Nevada (OSN), operated by the Instituto de Astrofísica de Andalucía (IAA-CSIC). 
This material is based upon work supported by the National Aeronautics and Space Administration under Agreement No.\ 80NSSC21K0593 for the program ``Alien Earths''.
The results reported herein benefited from collaborations and/or information exchange within NASA’s Nexus for Exoplanet System Science (NExSS) research coordination network sponsored by NASA’s Science Mission Directorate.
GD acknowledges funding from Magdalen College, Oxford.
The ULiege's contribution to SPECULOOS has received funding from the European Research Council under the European Union's Seventh Framework Programme (FP/2007-2013) (grant Agreement n$^\circ$ 336480/SPECULOOS), from the Balzan Prize and Francqui Foundations, from the Belgian Scientific Research Foundation (F.R.S.-FNRS; grant n$^\circ$ T.0109.20), from the University of Liege, and from the ARC grant for Concerted Research Actions financed by the Wallonia-Brussels Federation. MG and EJ are F.R.S-FNRS Senior Research Directors. 
This work is supported by a grant from the Simons Foundation (PI Queloz, grant number 327127).
J.d.W. and MIT gratefully acknowledge financial support from the Heising-Simons Foundation, Dr. and Mrs. Colin Masson and Dr. Peter A. Gilman for Artemis, the first telescope of the SPECULOOS network situated in Tenerife, Spain.
This work is supported by the Swiss National Science Foundation (PP00P2-163967, PP00P2-190080 and the National Centre for Competence in Research PlanetS).
This work has received fund from the European Research Council (ERC)
under the European Union's Horizon 2020 research and innovation
programme (grant agreement n$^\circ$ 803193/BEBOP), from the MERAC foundation, and from the Science and Technology Facilities Council (STFC; grant n$^\circ$ ST/S00193X/1). 
TRAPPIST is funded by the Belgian Fund for Scientific Research (Fond National de la Recherche Scientifique, FNRS) under the grant FRFC 2.5.594.09.F, with the participation of the Swiss National Science Fundation (SNF).
Visiting Astronomer at the Infrared Telescope Facility, which is operated by the University of Hawaii under contract 80HQTR24DA010 with the National Aeronautics and Space Administration.
This article is based on observations made in the 
Two-meter Twin Telescope (TTT\footnote{\url{http://ttt.iac.es}}) sited at the Teide Observatory of the Instituto 
de Astrofísica de Canarias (IAC), that Light Bridges operates in Tenerife, Canary Islands (Spain). The observation time rights (DTO) used for this research were consumed in the PEI 'SBSTLLAR25'.
This article used flash storage and GPU computing resources as Indefeasible Computer Rights (ICRs) being commissioned at the ASTRO POC project that Light Bridges will operate in the Island of Tenerife, Canary Islands (Spain). The ICRs used for this research were provided by Light Bridges in cooperation with Hewlett Packard Enterprise (HPE) and VAST DAT.
The data in this study were obtained with the T80 telescope at the Ankara University Astronomy and Space Sciences Research and Application Center (Kreiken Observatory) with the project number of 25C.T80.04.
The paper is based on observations made with the Kast spectrograph on the Shane 3m telescope at Lick Observatory. A major
upgrade of the Kast spectrograph was made possible through generous gifts from the Heising-Simons Foundation and William and
Marina Kast. We acknowledge that Lick Observatory sits on the unceded ancestral homelands of the Chochenyo and Tamyen Ohlone
peoples, including the Alson and Socostac tribes, who were the
original inhabitants of the area that includes Mt. Hamilton.
Some of the observations in this paper made use of the High-Resolution Imaging instrument Zorro and were obtained under Gemini LLP Proposal Number: GN/S-2021A-LP-105. Zorro was funded by the NASA Exoplanet Exploration Program and built at the NASA Ames Research Center by Steve B. Howell, Nic Scott, Elliott P. Horch, and Emmett Quigley. Zorro was mounted on the Gemini South telescope of the international Gemini Observatory, a program of NSF’s OIR Lab, which is managed by the Association of Universities for Research in Astronomy (AURA) under a cooperative agreement with the National Science Foundation. on behalf of the Gemini partnership: the National Science Foundation (United States), National Research Council (Canada), Agencia Nacional de Investigación y Desarrollo (Chile), Ministerio de Ciencia, Tecnología e Innovación (Argentina), Ministério da Ciência, Tecnologia, Inovações e Comunicações (Brazil), and Korea Astronomy and Space Science Institute (Republic of Korea).
This work makes use of observations from the LCOGT network. Part of the LCOGT telescope time was granted by NOIRLab through the Mid-Scale Innovations Program (MSIP). MSIP is funded by NSF.
This research has made use of the Exoplanet Follow-up Observation Program (ExoFOP; DOI: 10.26134/ExoFOP5) website, which is operated by the California Institute of Technology, under contract with the National Aeronautics and Space Administration under the Exoplanet Exploration Program.
Funding for the TESS mission is provided by NASA's Science Mission Directorate. KAC acknowledges support from the TESS mission via subaward s3449 from MIT.
We acknowledge the use of public TESS data from pipelines at the TESS Science Office and at the TESS Science Processing Operations Center.
Resources supporting this work were provided by the NASA High-End Computing (HEC) Program through the NASA Advanced Supercomputing (NAS) Division at Ames Research Center for the production of the SPOC data products.
J.d.W. and MIT gratefully acknowledge financial support from the Heising-Simons Foundation, Dr. and Mrs. Colin Masson and Dr. Peter A. Gilman for Artemis, the first telescope of the SPECULOOS network situated in Tenerife, Spain.
YGMC has been partially supported by UNAM-PAPIIT-IG101224. 

\section*{Data Availability}
The \emph{TESS} photometric observations that we used in this work are available via the Mikulski Archive for Space Telescopes (MAST) and the ExoFOP-\emph{TESS} platform. 
Our ground-based photometric time-series are also available via the ExoFOP-\emph{TESS} platform.



\bibliographystyle{mnras}
\bibliography{TOI-7166b_bib} 

\vspace{1cm}
\noindent 
$^1$Instituto de Astrofísica de Canarias (IAC), Calle Vía Láctea s/n, 38200, La Laguna, Tenerife, Spain\\
$^2$Astrobiology Research Unit, Université de Liège, Allée du 6 Août 19C, B-4000 Liège, Belgium\\
$^3$Department of Earth, Atmospheric and Planetary Science, Mas- sachusetts Institute of Technology, 77 Massachusetts Avenue, Cam- bridge, MA 02139, USA\\
$^4$Instituto de Astrofísica de Andalucía (IAA-CSIC), Glorieta de la Astronomía s/n, 18008 Granada, Spain\\
$^5$Department of Physics and Kavli Institute for Astrophysics and Space Research, Massachusetts Institute of Technology, Cambridge, MA 02139, USA\\
$^6$Department of Astronomy \& Astrophysics, UC San Diego, La Jolla, USA\\
$^7$School of Physics \& Astronomy, University of Birmingham, Edg- baston, Birmingham B15 2TT, UK\\
$^8$Departamento de Astrofísica, Universidad de La Laguna, Avda. As- trofísico Francisco Sánchez, 38206 La Laguna, Tenerife, Spain\\
$^9$Light Bridges, SL. Observatorio Astronómico del Teide\\
$^{10}$Department of Astronomy \& Space Sciences, Faculty of Science, Ankara University, TR-06100, Ankara, Türkiye\\
$^{11}$Ankara University, Astronomy and Space Sciences Research and Application Center (Kreiken Observatory), Incek Blvd., TR-06837, Ahlatlıbel, Ankara, Türkiye\\
$^{12}$Ankara University, Graduate School of Natural and Applied Sciences, Department of Astronomy and Space Sciences, Ankara, Türkiye\\
$^{13}$Department of Physics and Astronomy, Vanderbilt University, Nashville, TN 37235, USA\\
$^{14}$Center for Astrophysics | Harvard \& Smithsonian, 60 Garden Street, Cambridge, MA 02138, USA\\
$^{15}$Center for Space and Habitability, University of Bern, Gesellschaftsstrasse 6, 3012, Bern, Switzerland\\
$^{16}$Paris Region Fellow, Marie Sklodowska-Curie Action\\
$^{17}$AIM, CEA, CNRS, Université Paris-Saclay, Université de Paris, F-91191 Gif-sur-Yvette, France\\
$^{18}$Universidad Nacional Autónoma de México, Instituto de Astronomía, AP 70-264, CDMX 04510, México\\
$^{19}$Cavendish Laboratory, JJ Thomson Avenue, Cambridge CB3 0HE, UK\\
$^{20}$SUPA Physics and Astronomy, University of St. Andrews, Fife, KY16 9SS Scotland, UK\\
$^{21}$NASA Ames Research Center, Moffett Field, CA 94035, USA\\
$^{22}$Space Sciences, Technologies and Astrophysics Research (STAR) Institute, Université de Liège, Allée du 6 Août 19C, B-4000 Liège, Belgium\\
$^{23}$Bay Area Environmental Research Institute, Moffett Field, CA 94035, USA\\
$^{24}$Department of Astrophysics, University of Oxford, Denys Wilkinson Building, Keble Road, Oxford OX1 3RH, UK\\
$^{25}$Magdalen College, University of Oxford, Oxford OX1 4AU, UK\\
$^{26}$Institute for Particle Physics and Astrophysics, ETH Zürich, Wolfgang-Pauli-Strasse 2, 8093 Zürich, Switzeland\\
$^{27}$South African Astronomical Observatory, P.O. Box 9, Observatory, Cape Town 7935, South Africa\\
$^{28}$Hazelwood Observatory, Australia\\
$^{29}$SETI Institute, Mountain View, CA 94043 USA/NASA Ames Research Center, Moffett Field, CA 94035 USA\\


\appendix

\section{TOI-7166.01 observations log.}

In this appendix, we list the ground-based observations of TOI-7166\,b: Telescope, observation date, filter, exposure time, full width at half maximum (FWHM), photometric aperture, and detrended parameters that we use during our global analysis.

\begin{table*}
\caption{TOI-7166.01 observations log.}
 \begin{center}
 {\renewcommand{\arraystretch}{1.}
 \begin{tabular}{l l c c c c c cccccc}
 \toprule
Telescope & Date (UT) & Filter &  Exptime  &  FWHM & Aperture & comment & detrended parameter  \\ 
       &           &        &     [second] & [arcsec] & [arcsec]  & &   \\
 \hline
SPECULOOS-S-1.0m/Ganymede & June 8 2025 & Sloan-$z'$ & 13 & 3.8 & 2.0 & Full transit & Time + FWHM  \\
LCO-McD-1.0m & June 8 2025 & Sloan-$i'$ & 160 & 1.9 & 3.5 & Full transit & Time \\
SPECULOOS-S-1.0m/Europa & June 20 2025 & Sloan-$z'$ & 13 & 1.8 & 2.6  & Full transit & Time + FWHM  \\
SPECULOOS-S-1.0m/Io & June 20 2025 & Sloan-$g'$ & 140 & 2.8 & 2.7 & Full transit & Time + Sky  \\
TRAPPIST-S-0.6m & June 20 2025 & Rc & 140 & 2.5 & 4.4 & Full transit & Time + Airmass  \\
SPECULOOS-N-1.0m/Artemis & July 4 2025 & Sloan-$g'$ & 140 & 1.1 & 1.7  & Full transit & FWHM + Airmass \\
LCO-CTIO-1.0m & July 4 2025 & Sloan-$r'$ & 150 & 2.6 & 5.0 & Full transit & Time + dy \\
LCO-SAAO-1.0m & July 16 2025 & Sloan-$r'$ & 150 & 2.0 & 4.3 & Full transit & Time + FWHM \\
TTT3-2.0m & July 4 2025 & SDSSy & 60 & 1.1 & 2.1 & Full transit & Time + FWHM + dx \\
TTT1-0.8m & July 4 2025 & SDSSr, SDSSg & 90,90 & 1.1,1.3 & 2.2,2.3 &  not included & --\\ 
TTT3-2.0m & July 16 2025 & SDSSg & 60 & 1.4 & 1.8 & Full transit & Time + dy \\
TTT3-2.0m & July 29 2025 & SDSS gp & 60 & 1.9 & 2.9 & Full transit & Airmass + Time \\
SPECULOOS-N-1.0m/Artemis & July 29 2025 & $I+z$ & 13 & 1.2 & 2.3  & Full transit & Time \\
AUKR-T80 & July 29 2025 & Sloan-$i'$ & 100 & 2.4 & 4.8 & Full transit & FWHM + Airmass  \\
OSN-1.5m & July 29 2025 & Ic, V& 60,240 & 2.1, 2.8 & 4.6, 5.6 &  not included & -- \\
AUKR-T80 & Aug 24 2025 & Sloan-$i'$ & 120 & 3.1 & 6.8 & Full transit & FWHM + Airmass  \\
\hline
 \end{tabular}}
 \label{tab:GB_obs_table}
 \end{center}
\end{table*}

\section{Priors for the joint modeling of the transit light curves of TOI-7166.}

In this appendix, we present the parameter priors for the joint modeling of the transit light curves using the Metropolis-Hasting technique implemented in {\tt Trafit} code.

\begin{table*}
\caption{Priors for the joint modeling of the transit light curves of TOI-7166\,b. Normal priors are indicated as $\mathcal{N}$(mean, standard deviation) and uniform distribution are indicated as $\mathcal{U}$(lower bound, upper bound).} 
	\begin{center}
		{\renewcommand{\arraystretch}{1.3}
			\begin{tabular}{ll}
		\hline
				Parameter &  Value    \\   
	    \hline
            \multicolumn{2}{l}{\it Stellar parameters}  \\
        \hline
			 Quadratic Limb-darkening  $u_{\rm 1,TESS}$ &  $\mathcal{N}(0.31, 0.01)$   \\
			Quadratic Limb-darkening 	$u_{\rm 2,TESS}$ &  $\mathcal{N}(0.22, 0.05)$   \\
			Quadratic Limb-darkening 	$u_{\rm 1,{\rm Sloan-z'}}$ &  $\mathcal{N}(0.24, 0.02)$   \\
			Quadratic Limb-darkening 	$u_{\rm 2,{\rm Sloan-z'}}$ &  $\mathcal{N}(0.19, 0.05)$  \\
             Quadratic Limb-darkening     $u_{\rm 1,{\rm I+z'}}$ &  $\mathcal{N}(0.26, 0.02)$   \\
			 Quadratic Limb-darkening 	$u_{\rm 2,{\rm I+z'}}$ &  $\mathcal{N}(0.20, 0.04)$  \\
            Quadratic Limb-darkening     $u_{\rm 1,{\rm SDSSy}}$ &  $\mathcal{N}(0.22, 0.01)$   \\
			Quadratic Limb-darkening 	$u_{\rm 2,{\rm SDSSy}}$ &  $\mathcal{N}(0.17, 0.04)$  \\
			Quadratic Limb-darkening 	$u_{\rm 1,{\rm Sloan}-i'}$ &  $\mathcal{N}(0.345 0.019)$   \\
			Quadratic Limb-darkening 	$u_{\rm 2,{\rm Sloan}-i'}$ &  $\mathcal{N}(0.24, 0.067)$   \\
			Quadratic Limb-darkening 	$u_{\rm 1,{\rm Sloan}-r'}$ &  $\mathcal{N}(0.56, 0.033)$  \\
			Quadratic Limb-darkening 	$u_{\rm 2,{\rm Sloan}-r'}$ &  $\mathcal{N}(0.28, 0.085)$   \\
            Quadratic Limb-darkening      $u_{\rm 1,{\rm Sloan}-g'}$ &  $\mathcal{N}(0.56, 0.033)$  \\
			Quadratic Limb-darkening 	$u_{\rm 2,{\rm Sloan}-g'}$ &  $\mathcal{N}(0.28, 0.08)$   \\
            Quadratic Limb-darkening     $u_{\rm 1,{\rm Johnson}-Rc}$ &  $\mathcal{N}(0.38, 0.02)$  \\
			Quadratic Limb-darkening 	$u_{\rm 2,{\rm Johnson}-Rc}$ &  $\mathcal{N}(0.25, 0.04)$   \\
        	Effective temperature, $T_{\rm eff}$ [K]   &  $ \mathcal{N}(3100, 50)$  \\
			Surface gravity, $\log g_\star$ [dex]     &  $ \mathcal{N}( 5.02, 0.02 )$  \\
			Metallecity, $\mathrm{[Fe/H]}$ [dex]   &  $ \mathcal{N}( -0.20, 0.12) $ \\
			Stellar mass, $M_\star$  [$M_\odot$]    & $ \mathcal{N}(0.190 ,0.004) $  \\
			Stellar radius, $R_\star$  [$R_\odot$]     & $ \mathcal{N}( 0.220,  0.006)$ \\
        \hline
         \multicolumn{2}{l}{\it Planetary parameters}  \\
        \hline
        Orbital period, $P$ [days] & $\mathcal{U}(12.9,12.95)$ \\
        Impact parameters, $b$ & $\mathcal{U}(0.,0.8)$ \\
        Transit timing, $T_0$ [BJD-TDB] & $\mathcal{U}(2460847.7,2460847.8)$ \\
        Transit depth, $R^2_p/R^2_\star$ (ppt) & $\mathcal{U}(6,10)$ \\
        \hline
		\end{tabular}} 
	\end{center}
	\label{tab:priors}
\end{table*}

\section{TESS and ground-based transit light curves}
In this appendix, we show individual transit light curve observed from SPECULOOS-North/-South, TRAPPIST-South, LCOGT-1m0, TTT1, TTT3, OSN-1.5m and AUKR-T80 telescopes.

\begin{figure*}
	\centering
     \includegraphics[scale=0.3]{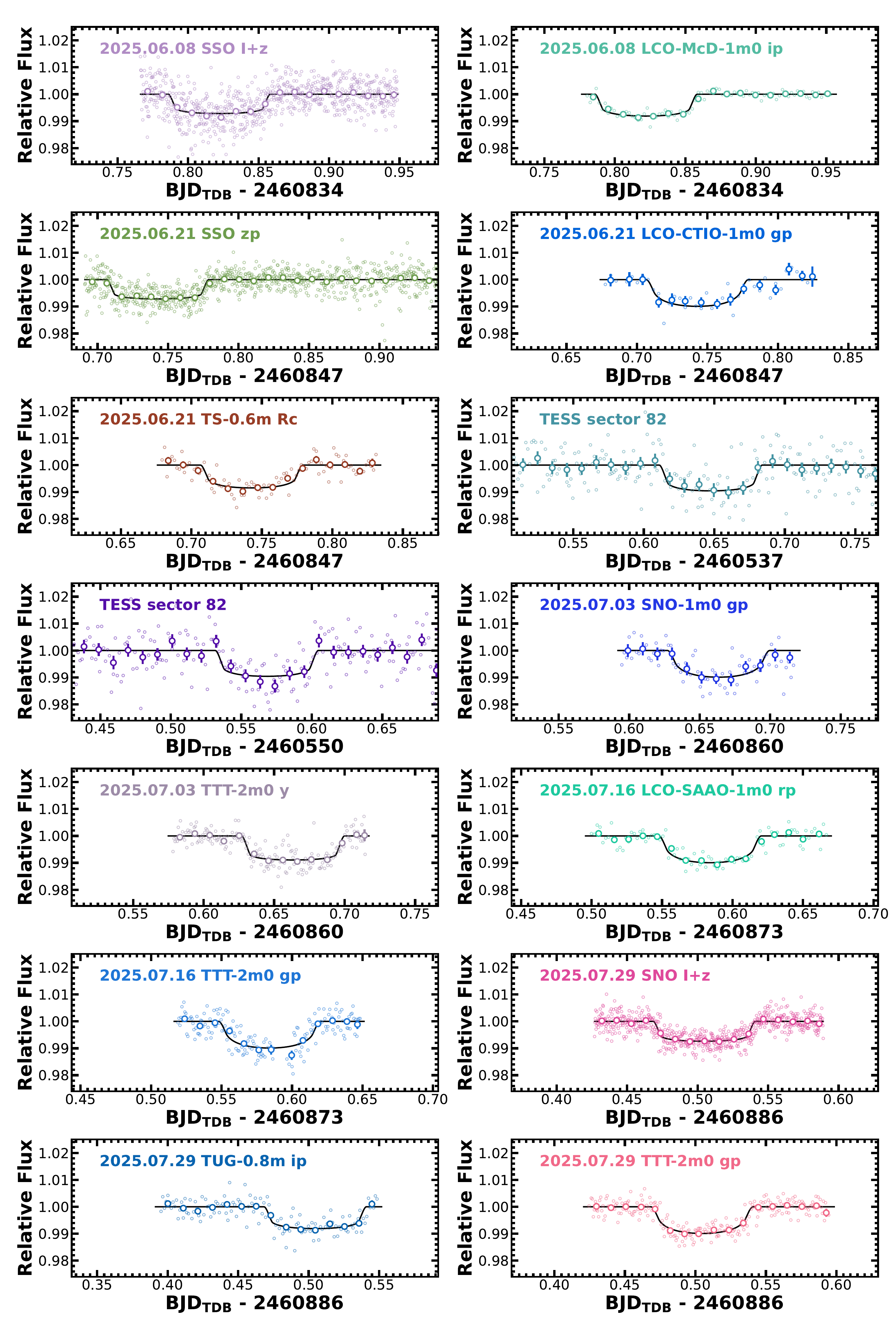} 
	\caption{\emph{TESS} and ground-based transit light curves for TOI-7166\,b. The colored data points show the relative flux and the black lines show the best-fitting transit model superimposed.} 
	\label{fig:TESS_GB_LCs_time}
\end{figure*}

\section{Transit fit posterior distributions.}

In this appendix, we show the corner diagram of the posterior probability distribution of the stellar and planetary physical parameters from our global MCMC analysis.

\begin{figure*}
	\centering
     \includegraphics[scale=0.3]{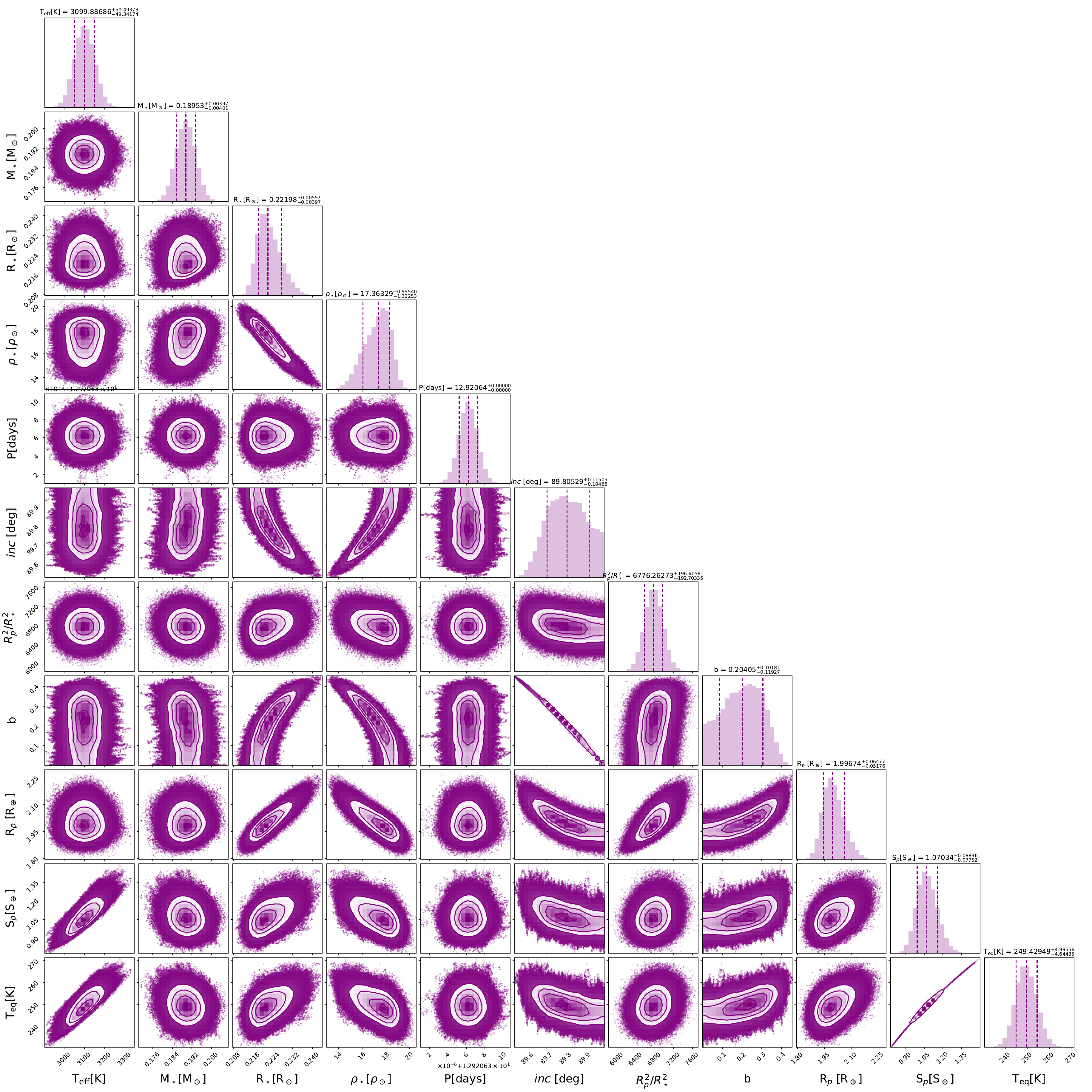} 
	\caption{Corner diagram of the posterior probability distribution from the our global analysis for the stellar and planetary parameters.} 
	\label{fig:Corner_plot}
\end{figure*}

\bsp	
\label{lastpage}
\end{document}